# Nine-hour X-ray quasi-periodic eruptions from a low-mass black hole galactic nucleus

G. Miniutti[1,*], R. D. Saxton[2], M. Giustini[1], K. D. Alexander[3,•], R.P Fender[4,5], I. Heywood[4,6,7], I. Monageng[8], M. Coriat[9], A. K. Tzioumis[10], A. M. Read[11], C. Knigge[12], P. Gandhi[12], M. L. Pretorius[8] and B. Agís-González[13]

[1] Centro de Astrobiología (CSIC-INTA), ESAC campus, 28692 Villanueva de la Cañada, Madrid, Spain

[2] Telespazio-Vega UK for ESA, European Space Astronomy Centre, Operations Department, Villanueva de la Cañada, Madrid, Spain

[3] Center for Interdisciplinary Exploration and Research in Astrophysics (CIERA) and Department of Physics and Astronomy, Northwestern University, Evanston, IL 60208, USA

[4] Astrophysics, University of Oxford, Denys Wilkinson Building, Keble Road, Oxford OX1 3RH, UK

[5] Department of Astronomy, University of Cape Town, Private Bag X3, Rondebosch 7701, South Africa

[6] Department of Physics and Electronics, Rhodes University, PO Box 94, Grahamstown, 6140, South Africa

[7] South African Radio Astronomy Observatory, 2 Fir Street, Observatory, 7925, South Africa

[8] South African Astronomical Observatory, PO Box 9, Observatory 7935, Cape Town, South Africa

[9] IRAP, Université de Toulouse, CNRS, CNES, UPS, Toulouse, France

[10] Australia Telescope National Facility, CSIRO, PO Box 76, Epping, New South Wales 1710, Australia

[11] Department of Physics and Astronomy, Leicester University, Leicester LE1 7RH, UK

[12] School of Physics & Astronomy, University of Southampton, Highfield, Southampton, SO17 1BJ, UK

[13] Institut d'Astrophysique et de Géophysique, Université de Liège, Allée du 6 Août 19c, 4000 Liège, Belgium

[*] gminiutti@cab.inta-csic.es

[•] NASA Einstein Fellow

**In the past two decades, high amplitude electromagnetic outbursts have been detected from dormant galaxies and often attributed to the tidal disruption of a star by the central black hole[1,2]. X-ray emission from the Seyfert 2 galaxy GSN 069 (2MASX J01190869-3411305) at redshift z = 0.018 was first detected in 2010 July and implies an X-ray brightening of more than a factor of 240 over ROSAT observations performed 16 years earlier[3,4]. The emission has smoothly decayed over time since 2010, possibly indicating a long-lived tidal disruption event[5]. The X-ray spectrum is ultra-soft and can be described by accretion disc emission with luminosity proportional to the fourth power of the disc temperature during long-term evolution. Here we report observations of X-ray quasi-periodic eruptions from the nucleus of GSN 069 over the course of 54 days, 2018 December onwards. During these eruptions, the X-ray count rate increases by up to two orders of magnitude with event duration of just over 1 hour and recurrence time of about 9 hours. These eruptions are associated with fast spectral transitions between a cold and a warm phase in the accretion flow around a low-mass black hole (of approximately $4 \times 10^5$ solar masses) with peak X-ray luminosity of $\sim 5 \times 10^{42}$ ergs per second. The warm phase has a temperature of about 120 electronvolts, reminiscent of the typical soft X-ray excess, an almost universal thermal-like feature in the X-ray spectra of luminous active nuclei[6,7,8]. If the observed properties are not unique to GSN 069, and assuming standard scaling of timescales with black hole mass and accretion properties, typical active galactic nuclei with more massive black holes can be expected to exhibit high-amplitude optical to X-ray variability on timescales as short as months or years[9].**

Since 2018 December 24, GSN 069 has exhibited peculiar high amplitude, short timescale X-ray variability, first detected during an XMM-Newton observation (XMM3, see Extended Data Table 1). The XMM-Newton light curve is characterized by two bright flares (bursts) with count rate increases by factors of $\sim 22$ and $\sim 31$ respectively in the 0.4-2 keV band (Figure 1a). The bursts are separated by $\sim 29.8$ ks, their profile is close to symmetric with similar rise and decay times of $\sim 1.8$ ks, and the total event duration is $\sim 4.5$ ks. This unexpected X-ray variability prompted us to request a longer XMM-Newton Director Discretionary Time (DDT) observation performed on 2019 January 16/17 (XMM4). Five bursts are detected with varying amplitudes, corresponding to count rate increases by factors of $\sim 19$ to $\sim 28$, and longer recurrence time ($\sim 32.15$ ks) than during XMM3 (Figure 1b). Finally, a Chandra DDT observation was performed on 2019 February 14/15 during which three further bursts are detected with count rate variations by factors of $\sim 13$ to $\sim 22$, and with recurrence time of $\sim 32.7$ ks (Figure 1c). We point out that no bursts were observed in a potentially long enough XMM-Newton exposure (83 ks) on 2014 December 5 (XMM2), i.e. 4 years before the XMM3 discovery observation.

The observed X-ray variability is characterized by short, high amplitude quasi-periodic X-ray bursts over a rather stable flux level (Figure 1). This type of variability has not hitherto been observed in active galactic nuclei (AGN). Hereafter, we refer to these new phenomena as X-ray *Quasi-Periodic Eruptions* (QPEs) to differentiate them from the gentler, quasi-sinusoidal modulation of the standard quasi-periodic oscillations (QPOs) that are often observed in X-ray binaries[10] and, more recently, in a handful of supermassive accreting black holes[11,12] (BHs).

Over the 54 days probed by our recent observations, the QPE amplitude decreases with time, and a simple linear extrapolation predicts no QPEs ~ 2019 late June onwards, if the current trend continues. The average QPE duty cycle is ~ 6 per cent, and it is well correlated with the average amplitude, while the QPE recurrence time seems to tend to a plateau at ~ 33 ks after an initial increase (Methods §IV and Extended Data Figure 4). Radio DDT observations with the MeerKAT, VLA, and ATCA radio facilities were carried out simultaneously with the Chandra observation (2019 February). GSN 069 is well detected at 1.3, 6, and 9 GHz with $L_{6\,GHz} \sim 1.9 \times 10^{36}$ erg s$^{-1}$ and a spectral index ~ - 0.7, consistent with optically thin synchrotron emission. No significant radio variability is observed in exposures simultaneous with X-ray QPEs (Methods §VI and Extended Data Figure 7).

Energy-selected light curves from the longest XMM4 observation folded on the average QPE recurrence time are shown in Figure 2, with two different choices of normalization to highlight the QPE amplitude (Figure 2a) and profile (Figure 2b) energy-dependence. QPEs measured in higher-energy bands i) peak earlier, ii) are narrower (shorter), and iii) have higher amplitudes than when measured at lower energies. The QPE energy-dependence is quantitatively displayed in Extended Data Figure 5. The maximum QPE amplitude ~ 100 is reached in the 0.6-0.8 keV band, to be compared with an amplitude of ~ 2 in the softest 0.2-0.3 keV band. The amplitude-energy relation implies very little variability below ~ 0.1 keV, indicating that QPEs are most likely restricted to the innermost accretion flow only.

The QPE spectral evolution during XMM4 is shown in Figure 3. During QPEs, the X-ray spectrum oscillates between a cold ~ 50 eV and a warm ~ 120 eV phase with peak X-ray luminosity of ~ $5 \times 10^{42}$ erg s$^{-1}$. The $L \sim T^4$ relation expected from constant-area disc emission is broken during QPEs, so that global mass accretion rate variation cannot explain the QPE spectral evolution. The variable thermal-like emission may be physically associated with Comptonization of the lower energy disc photons in a warm, optically thick corona or, e.g., with emission from different disc regions that are activated at different times during the cycle (Methods §V, Extended Data Table 3, and Extended Data Figure 6).

The QPE peak temperature (~ 120 eV) is remarkably similar to that of the standard AGN soft X-ray excess[6,7,8], which indicates that QPEs may be key to soft excess formation[13]. While in all XMM-Newton observations from 2010 December to 2019 January, the cold phase spectrum (i.e. the spectrum excluding QPE time intervals) is consistent with accretion disc emission, during the Chandra observation the cold phase has increased its temperature from ~ 50 eV to ~ 80 eV despite a lower soft X-ray luminosity, indicating the presence of a soft X-ray excess rather than pure disc emission (Extended Data Table 2 and Figure 3d). On the other hand, the temperature at the QPE peak remains at ~ 120 eV at the Chandra epoch. This suggests that we are possibly witnessing the QPE-driven formation of the soft X-ray excess in real time, with the Chandra cold phase spectrum representing an intermediate stage of soft-excess formation.

The discovery of QPEs in GSN 069 raises the question of whether this new phenomenon is unique to this specific source or more general. In this context, and in an attempt to identify other potential QPE candidates, it is interesting to spell out the properties of GSN 069 at the time when QPEs are first detected, namely: i) a small BH mass $M_{BH}$ ~ few × $10^5$ $M_\odot$; ii) a relatively high Eddington ratio of ~ 0.5; iii) a pure thermal disc spectrum with temperature significantly lower than that of the typical AGN soft X-ray excess; iv) an almost negligible X-ray power law component; v) the lack of any broad emission line in optical/UV spectra[4].

We have identified two other AGN with very similar properties, namely 2XMM J123103.2+110648[14] (hereafter J1231) and RX J1301.9+2747[15] (J1301) that are both characterized by low mass BHs, relatively high Eddington ratio, ultra-soft X-ray spectra with only weak power law tails, and no broad emission lines in their optical spectra. Remarkably, a ~ 3.8 hour QPO is detected in J1231[12,14], while J1301 exhibits high amplitude, possibly recurrent soft X-ray flares[15]. Exploring the possibility that these variability properties may be associated with QPEs could prove worthwhile. More generally, tidal disruption events (TDEs) exhibit, at least transiently, most of the key properties of GSN 069. For BH masses of ~ $10^6$-$10^7$ $M_\odot$, QPE recurrence times of the order of $10^5$-$10^6$ s are expected, and it would be valuable to search for X-ray variability on these timescales in the database of well monitored TDEs X-ray light curves. On the other hand, the typical AGN population is characterized by significantly higher BH masses than in GSN 069. The relatively long baselines (months/years) needed to detect two consecutive QPEs in these more massive systems may be the reason why no QPEs have been reported so far in AGNs.

The observed QPEs are reminiscent of the X-ray variability pattern of the black hole (BH) binaries GRS 1915+105 and IGR J17091-3624 in their *heartbeat* states[16,17,18]. It is worth mentioning that the heartbeat oscillations of both sources have been shown to be consistent with some version of the radiation-pressure disc instability[19,20], most likely modified by the presence of outflows[21,22]. As discussed in Methods §VII, a limit-cycle instability appears plausible in GSN 069 (but see the same Methods section for a series of possible alternative explanations). Assuming that the observed QPEs are driven by limit-cycle instabilities, we derive a mildly geometrically thick flow in GSN 069 even in the quiescent, cold state. As a consequence, the viscous diffusion (and thus the typical variability) timescale is significantly shorter than in razor-thin disc models. If the accretion flow parameters we derive in GSN 069 are not source-specific, our results may then provide a viable framework for interpreting the apparently unfeasibly rapid variability of some AGNs[23,24].

Some of the most extreme variability objects are the so-called (genuine, i.e. not absorption-induced) changing-look AGNs, i.e. objects in which high-amplitude continuum rise/decay is associated with the appearance/disappearance of the optical broad emission lines. Regardless of the QPE origin and physical interpretation, some changing-look AGNs may be naturally accounted for if they were experiencing QPEs extending down to the optical/UV band and to lower Eddington ratio than currently observed in GSN 069. Within this scenario, the observed QPE properties lead to predict i) similar numbers of rising and decaying changing-look AGNs due to the symmetric QPE profile, and ii) a developing/disappearing soft X-ray excess in the rise/decay phases.

We point out that the long-term variability of the two changing-look AGN Mrk 590[25,26,27] and Mrk 1018[23,28,29] is consistent with the latter prediction, as the soft X-ray excess has disappeared during decay (in both Mrk 590[26] and Mrk 1018[23]), and has recently re-emerged[27], together with the broad optical emission lines[30], in Mrk 590. If Mrk 590 is in the initial rise of a QPE, the strength of its soft excess should increase in the next years/decades. Detecting again a high state in the future would demonstrate the recurrent nature of the changing-look phenomenon in this AGN, strengthening the association with QPEs. We conclude that QPEs in massive AGN may possibly have been partially observed already as the rise/decay phases of (at least some) changing-look AGNs.

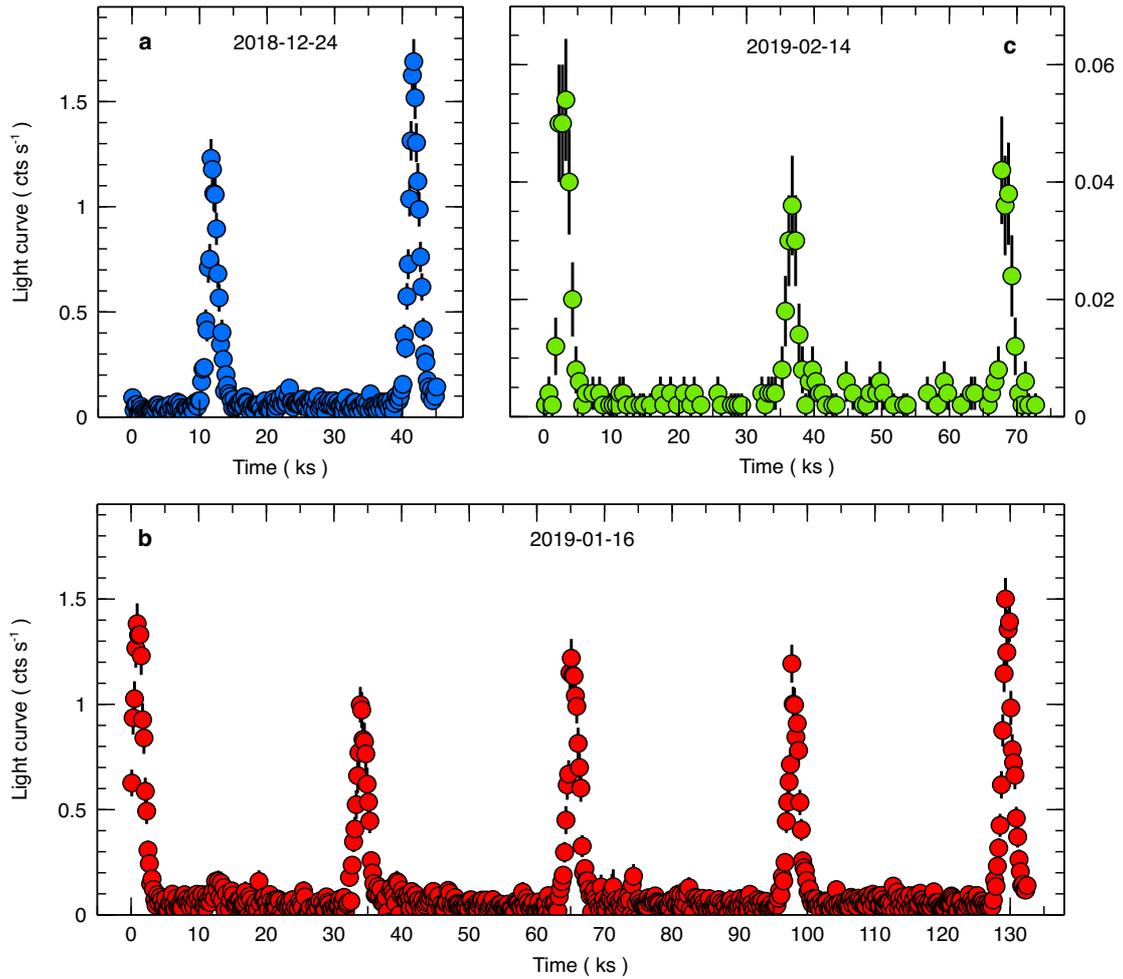

**Figure 1 | X-ray QPEs in XMM-Newton and Chandra observations, 2018 December onwards**
We show the background-subtracted 0.4-2 keV light curve from the XMM3 (**a**), XMM4 (**b**), and Chandra (**c**) observations. The x-axes are all on the same scale to highlight the similar QPE recurrence time over the 54 days spanned by the observations. We use time bins of 200 s for the XMM-Newton data, and of 500 s for the Chandra data. Note the different y-axis scale used for the Chandra data in **c**. Error bars represent 1-σ confidence intervals in all panels. Some of the error bars are smaller than the symbol size.

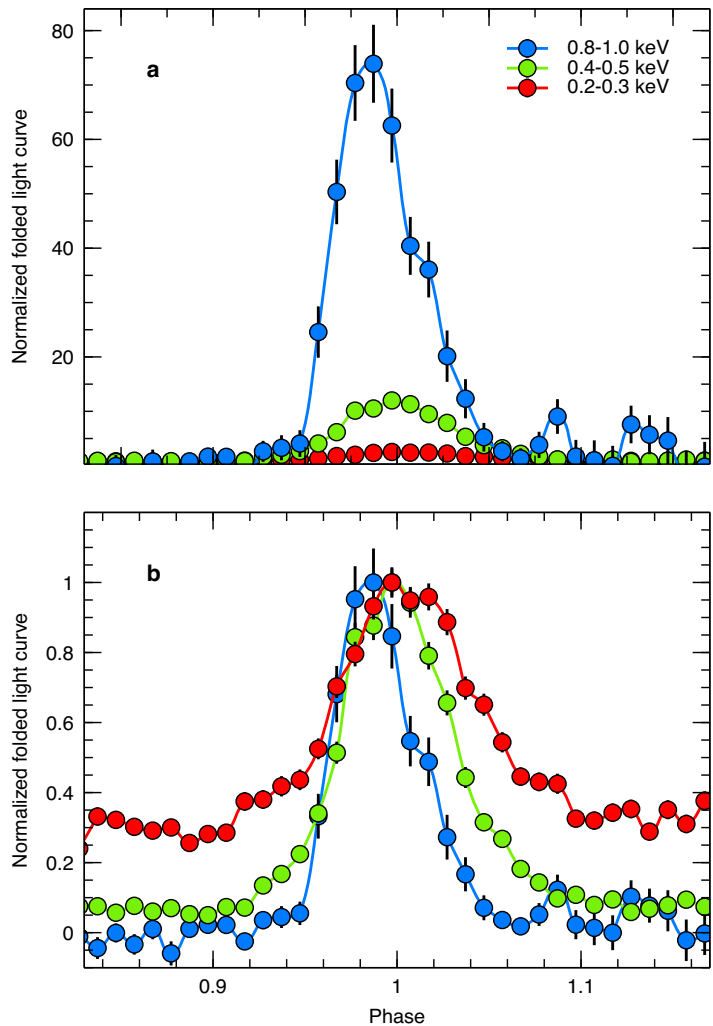

**Figure 2 | QPE energy-dependence from the longest XMM4 observation**

Folded light curves from the longest XMM4 observation around the QPE peak in three different energy bands and with two different normalizations: in **a**, light curves are normalized with respect to their quiescent level to highlight the QPE amplitude energy-dependence; in **b**, light curves are normalized with respect to their amplitude to highlight the different QPE peaking times and duration (width). On the x-axis, a phase interval $\Delta\varphi \sim 0.1$ corresponds to $\sim 3.2$ ks. Error bars represent the 1-σ confidence intervals. Some of the error bars are smaller than the symbol size.

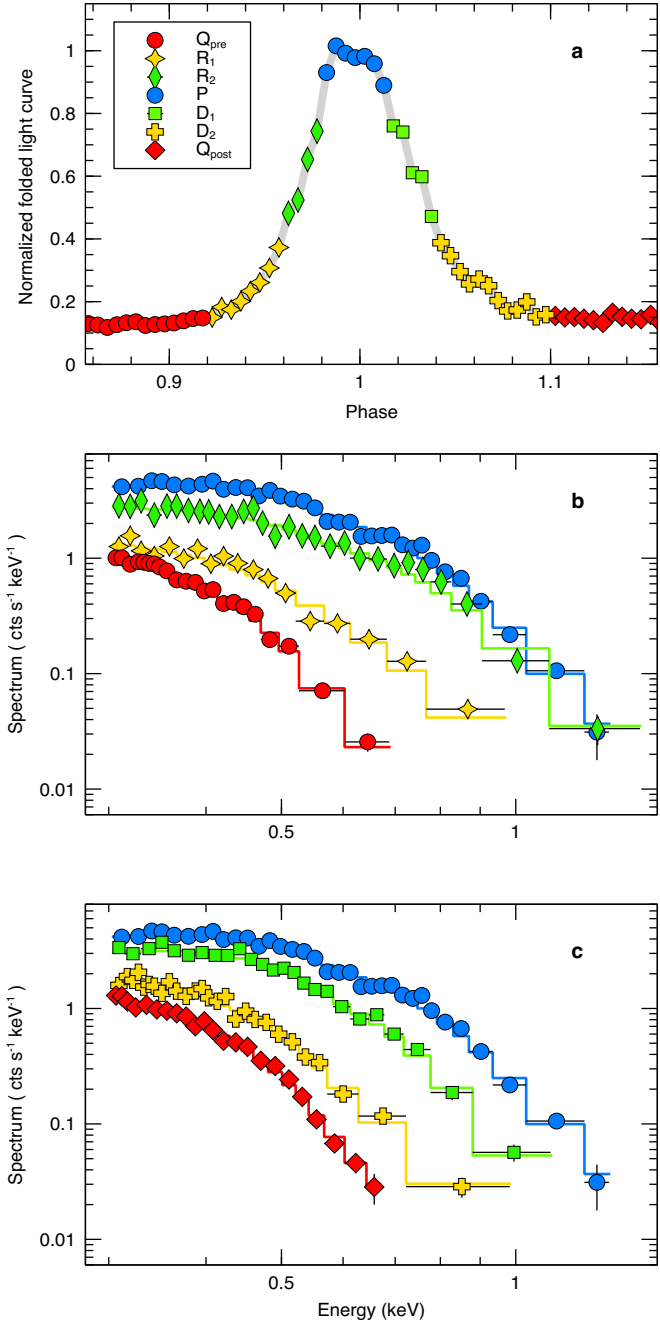

**Figure 3 | Phase-resolved X-ray spectra from the XMM4 observation**

In **a**, we show the XMM4 0.2-2 keV folded light curve identifying the phase-intervals used to extract phase-resolved spectra. In **b**, we show - from bottom to top - the $Q_{pre}$ to P spectra (i.e. from quiescence to peak) and best-fitting models (see Extended Data Table 3a); in **c**, we show - from top to bottom - the P to $Q_{post}$ spectra (i.e. from peak to quiescence) and best-fitting models. Error bars represent the 1-σ confidence intervals, although most error bars are smaller than the symbol size. $R_1$ and $R_2$ are two spectra during the QPE rise; $D_1$ and $D_2$ are two spectra during the decay.

# References

1. Komossa, S., Tidal disruption of stars by supermassive black holes: status of observations, *J. of High En. Astrphys.*, **7**, 148-157 (2015)

2. Auchettl, K., Guillochon, J. and Ramirez-Ruiz, E., New physical insights about Tidal Disruption Events from a comprehensive observational inventory at X-ray wavelengths, *Astrophys. J.*, **838**, 149 (2017)

3. Saxton, R. D., Read, A., Esquej, P., Miniutti, G. and Alvarez, E., Long-term AGN variability and the case of GSN 069, *Proc. of Science*, **126**, article id. 008, 1-12 (2011); preprint available at < https://pos.sissa.it/126/008/pdf>

4. Miniutti, G. et al., A high Eddington-ratio, true Seyfert 2 galaxy candidate: implications for broad-line region models. *Mon. Not. R. Astron. Soc.*, **433**, 1764-1777 (2013)

5. Shu, X. W. et al., A Long Decay of X-Ray Flux and Spectral Evolution in the supersoft active calactic nucleus GSN 069. *Astrophys. J. Letters*, **857**, L16 (2018)

6. Czerny, B., Nikołajuk, M., Różańska, A., Dumont, A.-M., Loska, Z. and Zycki, P. T., Universal spectral shape of high accretion rate AGN, *Astron. and Astrophys.*, **412**, 317-329 (2003)

7. Gierlínski, M. and Done, C., Is the soft excess in active galactic nuclei real ?, *Mon. Not. R. Astron. Soc.*, **349**, L7-L11 (2004)

8. Miniutti, G. et al., The XMM -Newton view of AGN with intermediate-mass black holes, *Mon. Not. R. Astron. Soc.*, **394**, 443-453 (2009)

9. MacLeod, C. L. et al., A systematic search for changing-look quasars in SDSS, *Mon. Not. R. Astron. Soc.*, **457**, 389-404 (2016)

10. van der Klis, M., Quasi periodic oscillations and noise in low-mass X-ray binaries, *Ann. Rev. of Astron. and Astrophys.*, **27**, 517-553 (1989)

11. Gierliński, M., Middleton, M., Ward, M. and Done, C., A periodicity of ~1hour in X-ray emission from the active galaxy RE J1034+396, *Nature*, **455**, 369-371 (2008)

12. Lin, D., Irwin, J. A., Godet, O., Webb, N. A. and Barret, D., A ~ 3.8 hr periodicity from an ultrasoft active galactic nucleus candidate, *Astrophys. J.*, **776**, L10 (2013)

13. Gronkiewicz, D. and Różańska, A., Warm and thick corona for magnetically supported disk in GBHB, preprint arXiv:1903.03641

14. Terashima, Y., Kamizasa, N., Awaki, H., Kubota, A. and Ueda, Y., A candidate active galactic nucleus with a pure soft thermal X-Ray spectrum, *Astrophys. J.*, **752**, 154 (2012)

15. Sun, L., Shu, X. W. and Wang, T. RX J1301.9+2747: A Highly Variable Seyfert Galaxy with Extremely Soft X-Ray Emission, *Astrophys. J.*, **768**, 67 (2013)

16. Belloni, T., Méndez, M., King, A. R., van der Klis, M. and van Paradijs, J., An Unstable Central Disk in the Superluminal Black Hole X-Ray Binary GRS 1915+105, *Astrophys. J.*, **479**, L145-L148 (1997)

17. Belloni, T., Méndez, M., King, A. R., van der Klis, M. and van Paradijs, J., A Unified Model for the Spectral Variability in GRS 1915+105, *Astrophys. J.*, **488**, L109-L112 (1997)

18. Altamirano, D. et al., The Faint "Heartbeats" of IGR J17091-3624: An Exceptional Black Hole Candidate, *Astrophys. J.*, **742**, L17 (2011)

19. Lightman, A. P. and Eardley, D. M., Black Holes in Binary Systems: Instability of Disk Accretion, *Astrophys. J.*, **187**, L1-L3 (1974)

20. Janiuk, A. and Czerny, B., On different types of instabilities in black hole accretion discs: implications for X-ray binaries and active galactic nuclei, *Mon. Not. R. Astron. Soc.*, **414**, 2186-2194 (2011)

21. Janiuk, A., Czerny, B. and Siemiginowska, A., Radiation Pressure Instability as a Variability Mechanism in the Microquasar GRS 1915+105, *Astrophys. J.*, **542**, L33-L36 (2000)

22. Janiuk, A., Grzedzielski, M., Capitanio, F. and Bianchi, S., Interplay between heartbeat oscillations and wind outflow in microquasar IGR J17091-3624, *Astron. and Astrophys.*, **574**, A92 (2015)

23. Noda, H. and Done, C., Explaining changing-look AGN with state transitions triggered by rapid mass accretion rate drop, *Mon. Not. R. Astron. Soc.*, **480**, 3898 (2018)

24. Dexter, J. and Begelman, M. C., Extreme AGN variability: evidence of magnetically elevated accretion ?, *Mon. Not. R. Astron. Soc.*, **483**, L17 (2019)

**25.** Denney, K. D. et al., The Typecasting of Active Galactic Nuclei: Mrk 590 no Longer Fits the Role, *Astrophys. J.*, **796**, 134 (2014)

**26.** Rivers, E., Markowitz, A., Duro, R. and Rothschild, R., A Suzaku Observation of Mkn 590 Reveals a Vanishing Soft Excess, *Astrophys. J.*, **759**, 63 (2012)

**27.** Mathur, S. et al., The Changing-look Quasar Mrk 590 Is Awakening, *Astrophys. J.*, **866**, 123 (2018)

**28.** Cohen, R. D., Rudy, R. J., Puetter, R.C., Ake, T.B. and Foltz, C. B., Variability of Markarian 1018: Seyfert 1.9 to Seyfert 1, *Astrophys. J.*, **311**, 135-141 (1986)

**29.** McElroy, R. E. et al, The Close AGN Reference Survey (CARS). Mrk 1018 returns to the shadows after 30 years as a Seyfert 1, *Astron. and Astrophys.*, **593**, L8 (2016)

**30.** Raimundo, S. I. et al., MUSE observations of a changing-look AGN - I. The reappearance of the broad emission lines, *Mon. Not. R. Astron. Soc.* **486**, 123-140 (2019)

## Acknowledgements

The scientific results reported in this work are based on observations obtained with XMM-Newton, an ESA science mission with instruments and contributions directly funded by ESA Member States and NASA, the Chandra X-ray Observatory, the NASA/ESA Hubble Space Telescope, the Australia Telescope Compact Array (ATCA), the Karl G. Jansky Very Large Array (VLA), and the South African MeerKAT radio telescope. The ATCA is part of the Australia Telescope National Facility which is funded by the Australian Government for operation as a National Facility managed by CSIRO. The NRAO operating the VLA is a facility of the National Science Foundation operated under cooperative agreement by Associated Universities, Inc. The MeerKAT telescope is operated by SARAO, which is a facility of the National Research Foundation, an agency of the Department of Science and Innovation. We also thank the Neil Gehrels Swift Observatory for supporting a long-term monitoring campaign of GSN 069 over the years.

We warmly thank Norbert Schartel, Belinda Wilkes, Jamie Stevens, Mark Claussen, and Fernando Camilo for approving XMM-Newton, Chandra, ATCA, VLA, and MeerKAT DDT observations, as well as the operation and scheduling teams of all involved observatories and facilities.

This research has been partially funded by the Spanish State Research Agency (AEI) Project No. ESP2017-87676-C5-1-R and No. MDM-2017-0737 Unidad de Excelencia "María de Maeztu"- Centro de Astrobiología (INTA-CSIC). GM and MG also acknowledge funding by the Spanish State Research Agency (AEI) Project No. ESP2017-86582-C4-1-R and No. ESP2015-65597-C4-1-R respectively.

KDA acknowledges support provided by NASA through NASA Hubble Fellowship grant HST-HF2-51403.001 awarded by the Space Telescope Science Institute, which is operated by the Association of Universities for Research in Astronomy, Inc., for NASA, under contract NAS5-26555. IH acknowledges support of the Oxford Hintze Centre for Astrophysical Surveys which is funded through generous support from the Hintze Family Charitable Foundation. PG acknowledges support from STFC and a UGC-UKIERI Phase 3 Thematic Partnership. BAG acknowledges support provided by the Fonds de la Recherche Scientifique - FNRS, Belgium, under grant No. 4.4501.19.

GM thanks Ari Laor for critically reading the initial manuscript as well as for suggesting the acronym *QPE: Quasi-Periodic Eruption* to describe the observed X-ray variability. GM and MG thank Agnieszka Janiuk, Mikołaj Grzędzielski, and Agata Różańska for useful discussions on the physics of disc instabilities and soft excess. GM wishes to dedicate this work to his father who passed away four years before the XMM3 discovery observation.

## Authors contribution

GM was PI of all XMM-Newton, HST, and Chandra proposals, led the organization of the 2019 February X-ray/radio campaign, performed most of the X-ray data analysis, and wrote the article. RDS discovered the source during the XMM-Newton slew and analysed the first XMM-Newton slew data together with AMR. RDS also performed part of the XMM-Newton EPIC and OM data analysis. MG performed part of the XMM-Newton and Chandra data analysis, and actively contributed to all intermediate results before submission. KDA was PI of the VLA DDT request and performed VLA radio data analysis. MLP led the MeerKAT DDT request, and IH, RPF, and IM, performed MeerKAT data analysis. RPF was PI of the ATCA DDT request. AKT observed GSN 069 with ATCA and MC performed ATCA data analysis. CK and PG were responsible for the analysis of the HST/STIS UV data. BAG was Co-I of the XMM3 discovery observation. All authors contributed to discussing data analysis and results, read the article since first draft, and helped improving it during the whole process.

## Data and code availability

Most data used in this work are public and available from the corresponding data archives. Some remaining proprietary data will be available immediately after the initial proprietary period expires. Data may be available from the corresponding author on reasonable request. All figures were made in Veusz, a Python-based scientific plotting package developed by Jeremy Sanders and freely available at <veusz.github.io>.

# Methods

Methods are organized as follows. In §I we discuss the most relevant data reduction aspects for the X-ray, radio, and UV observations used in this work. The source position and identification are briefly discussed in §II, whereas §III presents the long-term flux and spectral evolution of GSN 069 since first X-ray detection in 2010 July. Model-independent QPE properties are presented in §IV, and §V discusses results from phase-resolved spectroscopy throughout the QPE cycle. In §VI, results from the quasi-simultaneous X-ray/radio campaign in 2019 February are presented. Finally, §VII and §VIII are devoted to some of the possible interpretations of the QPE phenomenon, suggesting criteria for the selection of other potential QPE candidates, as well as discussing the broader implications of the observed QPEs.

## I. Observations and data reduction

Following first X-ray detection[3,4] on 2010 July 14 by an XMM-Newton slew[31], GSN 069 was monitored in the X-rays with a series of Neil Gehrels Swift, XMM-Newton, and Chandra observations. Swift data have been presented in previous works[4,5], and we are continuing our UV/X-ray monitoring of GSN 069 at least until 2019 September; we defer to future work the presentation of the full monitoring campaign. Here we focus on the highest quality XMM-Newton and Chandra X-ray data from 2010 December (~ 5 months after first X-ray detection) to 2019 February. XMM-Newton observations in 2014 December and 2018 December were quasi-simultaneous with Hubble Space Telescope (HST) spectroscopic observations in the UV by the Space Telescope Imaging Spectrograph (STIS) using the Multi-Anode Microchannel Array (MAMA) detector with the G140L and G230L gratings (~ 1150-3180 Å). The Chandra observation (2019 February) was simultaneous with a series of radio observations performed with the Australia Telescope Compact Array (ATCA), the South-African MeerKAT radio telescope, and the Karl G. Jansky Very Large Array (VLA). A summary of the observations used in this work is presented in Extended Data Table 1.

**X-ray data reduction**

XMM-Newton and Chandra data were reduced as standard using the latest available versions of the dedicated XMMSAS and CIAO software packages respectively. As for XMM-Newton, in this work we use data from the XMM-Newton EPIC-pn camera only, due to its superior sensitivity; however, products have been extracted for the two MOS detectors as well, and results reported here have been double-checked against the MOS data. We have also reduced and analysed data from the Reflection Grating

Spectrometer - RGS[a] - and Optical Monitor (OM) on board XMM-Newton using standard XMMSAS routines. For the EPIC-pn and the Chandra ACIS-S data, X-ray source and background products were extracted from circular regions centred on the source and from off-source regions on the same detector chip respectively. All light curves were background-subtracted, as well as corrected for various effects (e.g. vignetting, dead-time, and bad pixels) using the tasks *epiclccorrr* (XMM-Newton) and *dmextract* (Chandra). Observation-dependent response and ancillary files for spectral products were generated during product extraction. For spectral and timing analysis we used the XSPEC and XRONOS software that are included in the HEAsoft distribution. The Chandra source position has been obtained using the *celldetect* CIAO task (§II).

**Radio data reduction**

Results from the ATCA, VLA, and MeerKAT observations are discussed in §VI. Here we report the most relevant aspects of the radio data reduction.

GSN 069 was observed with ATCA on 2019 January 26 (ATCA1) and February 14/15 (ATCA2). The total on-source time is 5.2 hrs for ATCA1 and 10.5 hrs for ATCA2. The telescope was in the highly compact H75 configuration with 10 baseline length below 90 m and 5 baselines ranging from 4.3 to 4.4 km. We observe simultaneously at 5.5 GHz and 9 GHz. Each frequency band was composed of 2048 × 1-MHz channels. We used PKS 1934-638 for bandpass and absolute flux calibration and 0132-389 to calibrate the per-antenna complex gains as a function of time. Flagging and initial calibration were carried out with the Multichannel Image Reconstruction, Image Analysis and Display (MIRIAD) software[32]. Once the complex gains were calibrated using external calibrator sources in MIRIAD, we converted the calibrated visibilities of the target field into measurement sets for further analysis with the Common Astronomy Software Application[33] (CASA). We applied a 8:1 frequency averaging to make the data volume more manageable. No time averaging was applied, so the standard 10 s integration time was preserved. Due to the inhomogeneous uv-coverage produced by the array configuration we used a uv-taper of 4″ and 7″ for deconvolving and imaging the 9 GHz and 5.5 GHz respectively. To search for potential variability in the radio band, we split each dataset into ~ 1 hr portions and imaged them, but we did not detect the source in any of these images.

We observed GSN 069 with the VLA for a total on-source time of ~ 3.7 hours between 2019 February 14 and 15. Observations with the VLA were conducted at a mean frequency of 6.0 GHz with GSN 069 was observed with MeerKAT on 14 and 15 February 2019 using 59 of the 64 antennas in the array, for a total on-source observing time of ~ 3 hours per epoch. Standard procedures were used within the CASA package to remove radio frequency interference from the data, and set the instrumental delays, bandpass and absolute flux scale via a 5 minute scan of the standard southern-sky calibrator source PKS J0408-6465. Time-dependent instrumental gain corrections were derived from an observation of the nearby secondary calibrator source J0155-4048. An imaging and phase-only self-calibration cycle was performed using *wsclean*[35] and *CubiCal*[36]. Following this, a direction-dependent calibration scheme was applied using a facet-based approach[37]. This was necessary to mitigate image corruptions introduced by the stronger off axis sources, in particular QSO B0115-342, located 22.5 arcminutes away from GSN 609. These effects are the cause of the increased effective noise in epoch 2. In order to nullify any systematic flux scale biases in the detection of any variability from GSN 069, the image from the second epoch was bootstrapped to the flux scale of the first using measurements of 542 compact field sources.

---

[a] The relatively low signal-to-noise of the RGS spectra prevents us from reporting any significant results. We only mention here that hints for an ionized absorber are seen in XMM1, confirming results from CCD-resolution EPIC spectral analysis (§III). No statistically significant spectral features appear to be associated with QPEs in RGS intensity-selected spectra during XMM3 and XMM4.

4 GHz of bandwidth while the VLA was transitioning from C configuration to B configuration. We used 3C147 as the flux and bandpass calibrator and J0153-3310 as the phase calibrator. The data were reduced and imaged using standard CASA routines. We used the *imtool* program[b] to fit a point source model in the image plane. Radio variability analysis was performed by using the *dftphotom* routine in the *pwkit* package. This routine uses discrete Fourier transforms to fit the visibilities with a point source model centred at the source coordinates, without the need for imaging. We also reduced and analysed archival data from an observation performed on 2 November 2017 (program 17B-027; PI: Shu) using a similar frequency setup (2 GHz total bandwidth with a mean frequency of 6 GHz) in B configuration. We reduced and imaged this data using the same technique described above. The time on source was only 4.5 minutes; we therefore could not test for variability within the duration of the 2017 observation.

**UV data reduction**

GSN 069 was observed twice spectroscopically in the far- and near-ultraviolet with the HST/STIS MAMA detector using the G230L and G140L gratings. All spectra were taken with the 52″×0.5″ slit centred on the core of the galaxy. The first observation was quasi-simultaneous with XMM2, the second with XMM3. In each epoch, 2×49 minutes far-ultraviolet, and 1×39 minutes near-ultraviolet exposures were obtained. We used the pipeline-calibrated 1-D spectra that are provided within the standard package of data products for our analysis. After verifying that there was no significant variability between the two far-ultraviolet spectra obtained in each epoch, we averaged them to yield a single far-ultraviolet spectrum for each epoch. Each of these was then combined with the corresponding near-ultraviolet spectrum by splicing them at 1715 Å to produce spectra covering the full ultraviolet range from ~1130 to 3100 Å. We finally shifted the wavelength scale to the rest frame, and the resulting HST UV spectra are shown in Extended Data Figure 2a.

---

[b] This is part of the *pwkit* package[34], available on line at <https://github.com/pkgw/pwkit>.

## II. Source position and identification

The Chandra X-ray position is RA 01:19:08.668; DEC -34:11:30.53, with a statistical uncertainty of less than 0.1 arcseconds (no boresight correction was applied). The VLA radio position is RA 01:19:08.658 (± 0.12 arcseconds); DEC -34:11:30.53 (± 0.42 arcseconds), consistent with the Chandra one. The 2MASS position of the galactic nucleus is RA 01:19:08.663 (± 0.16 arcseconds); DEC -34:11:30.52 (± 0.14 arcseconds), again consistent with the Chandra and VLA position. Extended Data Figure 1 shows a relatively wide 12′×12′ field of view from the Digitized Sky Survey (Extended Data Figure 1a) and a zoom into the central 1.7′×1.7′ region as imaged by the VLA and Chandra (Extended Data Figures 1b and 1c). The Chandra X-ray image of the innermost 12″×12″ region is shown in Extended Data Figure 1d, together with the VLA 6 GHz contours and the 2MASS position of GSN 069 as reference.

At the galaxy distance, the highest measured X-ray luminosity of $L_X \sim 1.1 \times 10^{43}$ erg s$^{-1}$ (see §III) is much higher than that of the most luminous Ultra-Luminous X-ray sources detected thus far, including HLX-1 in ESO 243-49[38]. The Galactic latitude of GSN 069 is -80.76 degrees which makes a foreground Galactic object highly unlikely. Moreover, assuming a distance of (say) 5 kpc, the X-ray luminosity would be $< 5 \times 10^{34}$ erg s$^{-1}$ at all epochs, orders of magnitude too low for a Galactic stellar-mass BH or a neutron star in outburst. The star-like UV spectrum, and the lack of optical/UV variability make a white dwarf system highly unlikely (see Extended Data Figure 2). In summary, the extragalactic nature of the X-ray source and its association with the Seyfert 2 nucleus of GSN 069 appear robust.

## III. Long-term evolution

The long-term properties studied so far are consistent with the current 2010-2019 outburst of GSN 069 being associated with either the tidal disruption of a star by the central BH, or with AGN re-activation[4,5]. Note that, in the former scenario, the long-lived X-ray emission may point towards the disruption of an evolved giant star[39]. AGN re-activation is instead suggested by the Seyfert 2 nature of GSN 069, as inferred from optical spectra taken with the Anglo Australian Telescope both before (2001 and 2003) and after (2010 October) first X-ray detection. We note that the historical bolometric luminosity[4], as estimated from the [O$_{III}$] luminosity, is $L_{hist} \sim 3\text{-}4 \times 10^{42}$ erg s$^{-1}$, i.e. about one order of magnitude lower than that at first X-ray detection (see §III below). Since $L_{hist}$ is representative of the averaged nuclear luminosity over the long timescales probed by the narrow line region, and considering the lack of X-ray detection[3,4] by ROSAT in 1994, this points towards past low level or intermittent nuclear activity (e.g. recurrent AGN outbursts or a series of TDEs).

In order to study the long-term spectral evolution of GSN 069, we consider the high-quality XMM-Newton and Chandra data, excluding time-intervals when QPEs are present. UV photometry by the Optical Monitor (OM) on board XMM-Newton shows no sign of long-term variability: in particular, the flux density in the bluest filter UVW2 (~212 nm) is consistent with being the same in XMM2 and XMM3 despite an X-ray flux variability by a factor of ~ 2. No UV variability is observed on short timescales during XMM4, despite the high-amplitude X-ray QPEs (Extended Data Figure 2c). HST/STIS spectra, quasi-simultaneous with XMM2 and XMM3, show very little variability (if any) down to ~ 120 nm (Extended Data Figure 2a). In fact, the UV spectrum strongly resembles that of intermediate-type main sequence stars, indicating the likely presence of a relatively young stellar nuclear cluster dominating the UV emission, as shown by the comparison of the STIS spectrum with that of type-B stars from the Pickles Atlas[40] (Extended Data Figure 2b). We then focus here on the X-ray data only. X-ray spectral fits are performed using $\chi^2$ minimisation, and we report errors at the 90 per cent confidence level.

The XMM-Newton (Chandra) spectra are grouped to a minimum of 50 (20) counts per energy bin and are considered down to 0.3 keV (0.4 keV), as calibration uncertainties affect lower energy data. The X-ray spectrum is ultra-soft and appears to be dominated by thermal disc emission with a weak hard X-ray tail (Extended Data Figure 3a). We consider a spectral model comprising a thermal accretion disc (the *diskbb*[41] model in XSPEC), a power law component, and Galactic (z=0) absorption (*tbabs*[42] in XSPEC). All spectra are fitted simultaneously. As the data quality above 1-2 keV is rather poor, the power law photon index $\Gamma$ is forced to be the same in all observations. Initial fits show that $\Gamma$ can take any value between 1.1 and 2.5, and we fix $\Gamma = 1.8$ in our analysis, which is a typical spectral shape for AGNs[43]. We reach a good description of the overall data set with $\chi^2 = 301$ for 256 degrees of freedom (hereafter dof), and the most relevant best-fitting parameters are reported in Extended Data Table 2. A relatively low column density ($N_H \sim 5 \times 10^{21}$ cm$^{-2}$) ionized absorber (log $\xi \sim 0.35$) is detected in XMM1 ($\Delta\chi^2 = -18$ for 2 dof) using the *zxipcf* model[44], as reported in previous analysis[4]. The absorber is not significant in any of the remaining observations, and it is therefore only included in the spectral model for XMM1. The (z=0) neutral column density is ~ $5.5 \times 10^{20}$ cm$^{-2}$, about twice the Galactic line of sight measurements from the LAB map[45] (~ $2.5 \times 10^{20}$ cm$^{-2}$), which may indicate some degree of intrinsic absorption.

Our best-fitting model shows that the disc temperature smoothly decreases from kT ~ 63 eV in XMM1 to ~ 47 eV in XMM4. On the other hand, the Chandra spectrum is characterized by a much higher temperature of ~ 82 eV. We measure a maximum X-ray luminosity $L_X$ ~ 1.1 × $10^{43}$ erg s$^{-1}$ in the 0.2-2 keV band during the XMM1 observation, i.e. ~ 5 months after first detection in 2010 July, corresponding to a bolometric luminosity of $L_{Bol}$ ~ 4.8 × $10^{43}$ erg s$^{-1}$. The best-fitting spectral energy distribution (SED) evolution is shown in Extended Data Figure 3b. Using E(B-V) = 1.7 × $10^{-22}$ × $N_H$ where $N_H$ = 5.5 × $10^{20}$ cm$^{-2}$, as derived from X-ray spectral analysis, the extrapolation of our best-fitting model during XMM2 and XMM3 (quasi-simultaneous with HST observations) implies that the active nucleus contribution in the HST band is of the order of ~ 30-40 per cent.

The historical evolution of the 0.3-2 keV X-ray flux since first X-ray detection in 2010 July is shown in Extended Data Figure 3c. The overall decay can be described either with an exponential law with e-folding timescale of ~ 5 yr and starting time $t_0$ ~3-4 years before first detection, or with a power law decay. In the latter case, the power law index and starting time cannot be simultaneously constrained, so that we fix the index to -5/3, i.e. to the expected value for TDEs[46,47], although other indices may be observed[2]: under this assumption, we obtain $t_0$ ~ 5 years before first detection. The statistical quality of the fits is poor and very similar between the two models (reduced $\chi^2$ ~ 15). We conclude that our ignorance of the light curve shape during the ~ 16 years between the latest ROSAT upper limit and first X-ray detection in 2010 July does not allow us to strongly prefer one decay law over the other. Regular long-term monitoring of GSN 069 may provide better constraints in the future. Note that the Chandra data point appears to be inconsistent with the overall decay and may be instead representative of an excursion into a low flux state. Similar short-term X-ray variability is also seen in previous Swift monitoring data[4,5]. Integrating the historical light curve, and assuming a typical accretion efficiency of ~ 6 per cent, we estimate an accreted mass of $M_{accr}$ ~ $10^{-2}$ $M_\odot$ since first detection, not inconsistent with a long-lived outburst from the (possibly partial) disruption of a star by the central BH.

In Extended Data Figure 3d, we show the 0.2-2 keV unabsorbed disc luminosity $L_{diskbb}$ as a function of the best-fitting disc temperature. A fit using the XMM-Newton data only results in $L_{diskbb}$ ~ $T^{4.5 \pm 0.5}$, consistent with the Planckian L ~ $T^4$ expectation for constant-area blackbody emission. We are then observing directly the accretion disc emission in the XMM1 to XMM4 observations. The Chandra data point is inconsistent with a disc blackbody origin as a rather abrupt temperature increase is associated with a drop in luminosity over the course of ~ 1 month. This suggests that, during the Chandra observation, the pure thermal disc is not dominating the soft X-rays anymore, indicating the presence of a much hotter thermal-like component strongly resembling the typical AGN soft X-ray excess, a component whose shape is similar to a blackbody with universal temperature kT ~ 100-200 eV and that is almost ubiquitous in the X-ray spectra of unobscured, radiatively efficient AGNs[6,7,8]. The fact that we are effectively fitting a soft X-ray excess rather than pure disc emission at the Chandra epoch suggests to take the estimated bolometric luminosity (and hence Eddington ratio) with caution (see also Extended data Table 2). In fact, assuming a two-component model for the Chandra data comprising a cold disc forced to obey to the $L \sim T^4$ relation (as in all other observations) plus a $\sim 80$ eV soft excess shows that $L_{Bol}$ ($L_{Bol} / L_{Edd}$) can be as high as $\sim 1.1 \times 10^{43}$ erg s$^{-1}$ ($\sim 0.22$). We conclude that, at the Chandra epoch, $L_{Bol} \sim 0.09$-$1.1 \times 10^{43}$ erg s$^{-1}$, and $L_{Bol} / L_{Edd} \sim 0.02$-$0.22$.

As the X-ray spectrum during the XMM-Newton observations is consistent with pure blackbody emission, the derived best-fitting parameters can be used to estimate the inner disc radius $R_{in}$ (and hence the BH mass $M_{BH}$). Following previous work[48], and assuming an intermediate inclination of 45 degrees w.r.t. the disc normal, we derive $R_{in} \sim 4 \times 10^{11}$ cm. Identifying $R_{in}$ with the innermost stable circular orbit (ISCO) for a non-rotating Schwarzschild BH ($R_{isco} = 6\, R_g = 6\, GM_{BH}\, c^{-2}$), we estimate $M_{BH} \sim 4 \times 10^5\, M_\odot$. Considering that $L_{Bol} \sim 4.8 \times 10^{43}$ erg s$^{-1}$ (see Extended Data Table 2), we infer an Eddington ratio $L_{Bol} / L_{Edd} \sim 0.95$ during XMM1. We estimate a factor of a few uncertainty in the derived $M_{BH}$ (and therefore Eddington ratio) mostly depending on the actual BH spin and observer inclination.

The power law emission above 2 keV is always weak with an almost epoch-independent luminosity of $\sim 1.1 \times 10^{40}$ erg s$^{-1}$ in the 2-10 keV band. Hence, the power law fractional contribution to the overall radiative output significantly increases as the overall luminosity drops, in line with the general correlation between X-ray bolometric correction and Eddington ratio in AGNs[49]. GSN 069 is however an extreme example of X-ray weakness above 2 keV at all epochs, with a 2-10 keV bolometric correction as high as $\sim 4000$ in XMM1 (2010 December). Constant hard X-ray luminosity is unusual in AGN (and even more so for low BH mass nuclei[8]) and may imply that this emission component is not associated with a compact optically thin X-ray corona as generally assumed, but rather with emission from diffuse hot gas and/or X-ray binaries in the host galaxy. Using the relationship between star-formation rate (SFR) and 2-10 keV luminosity[50], a SFR of $\sim 2\, M_\odot$ yr$^{-1}$ is needed to account for the observed 2-10 keV luminosity. The SFR in GSN 069 can be estimated from the reddening-corrected UV luminosity[51] as SFR [$M_\odot$ yr$^{-1}$] $= 1.4 \times 10^{-28}\, L_\nu$ (UV) [erg s$^{-1}$ Hz$^{-1}$]. From the HST spectroscopic observations, we derive $L_\nu$ (UV) $\sim 1.2 \times 10^{27}$ erg s$^{-1}$ Hz$^{-1}$ at $\sim 1500$ Å, where we have assumed the reddening associated with the measured X-ray column density[c]. From the estimated $L_\nu$ (UV) we then derive SFR $\lesssim 0.2\, M_\odot$ yr$^{-1}$, where the upper limit is motivated by the fact that we ascribe the whole UV luminosity to stellar processes, ignoring the likely 30-40 per cent contribution of the nucleus in the HST band. The derived

---

[c] This is higher than the Galactic value, but we are here interested in deriving a SFR upper limit, rather than an accurate value.

SFR upper limit is about one order of magnitude too low to account for the observed 2-10 keV X-ray luminosity. Using the 1.3 GHz luminosity (see §VI) as tracer of star-formation[50], an even lower SFR is obtained. We point out that a single Ultra-Luminous X-ray source in the galaxy could account for the whole observed 2-10 keV X-ray luminosity. However, in this case, the lack of hard X-ray variability over the ~ 8.5 years probed by our observations would remain puzzling.

## IV. QPE model-independent properties

We consider light curves from the XMM3, XMM4 and Chandra observations in the common 0.4-2 keV band (see Figure 1). All light curves are described with a simple model comprising an observation-dependent constant C representing the quiescent level, and a series of Gaussian functions with normalization N and width σ describing the QPEs. We define the QPE recurrence time $T_{rec}$ as the peak-to-peak time-interval between subsequent eruptions, and the QPE duration $T_{dur}$ as the FWHM of the best-fitting Gaussian model. The duty cycle is defined as $\Delta = T_{dur} / T_{rec}$, while the QPE amplitude is A=N/C.

The long-term QPE evolution is shown in Extended Data Figure 4, and has already been discussed earlier. In order to gain insights on the QPE energy-dependence we study energy-selected folded light curves from the longest XMM4 observation. As shown in Figure 2, the QPE amplitude, duration, and peaking time are all strongly energy-dependent. The amplitude energy-dependence is shown in Extended Data Figure 5a. The QPE amplitude A is much higher when measured at higher X-ray energies, with A ~ 93 in the 0.6-0.8 keV band to be compared with A ~ 2 in the 0.2-0.3 keV band. Extrapolation to lower energies implies that little/no variability is expected below 0.1 keV. The amplitude energy-dependence can be split into the energy-dependence of the quiescent count rate C=C(E) and of the Gaussian QPE normalization N=N(E). This is shown in Extended Data Figure 5b where C(E) and N(E) are normalized to the detector effective area, so that we effectively show the photon spectrum of the quiescent and QPE peak levels. The quiescent spectrum is consistent with the high-energy tail of a few tens of eV thermal model, while the peak spectrum is harder (warmer), suggesting that GSN 069 oscillates between a cold and a warm phase during QPEs. Extended Data Figure 5b also clarifies the reason behind the QPE amplitude energy-dependence: the different spectral shape of the two phases results into a much higher QPE amplitude (A=N/C) at higher energies, where C(E) and N(E) differ the most.

As already clear from Figure 2, QPEs in higher energy bands evolve faster and peak earlier than when measured in lower energy bands. The QPE duration $T_{dur}$ increases with energy from ~ 1,590 s in the 0.8-1 keV band to ~ 2,475 s in the 0.2-0.3 keV band, as shown in Extended Data Figure 5c. The QPE peaking time in the hardest 0.8-1 keV band leads all other energies with increasing lags up to ~ 510 s (Extended Data Figure 5d). The $T_{dur}$ and peaking time energy-dependence suggests that QPEs first develop in the innermost (hotter) regions of the accretion flow and propagate outwards. The faster evolution of QPEs at higher energies could then reflect the shorter accretion flow timescales of the inner, hotter w.r.t. outer, colder regions. Fitting simple linear relations, the QPE duration $T_{dur}$ and peak time delay $\Delta T_{peak}$ energy-dependence are consistent with

$T_{dur} = (2806 \pm 66) - (1325 \pm 115) \times E$ [s];

$\Delta T_{peak} = (723 \pm 30) - (819 \pm 55) \times E$ [s],

where E is in units of keV and the delay is computed with respect to the 0.8-1 keV band (which is assumed to have zero delay). All results discussed above are based on the averaged QPE properties during the XMM4 observation, as inferred from folded light curves. We defer to a future work the detailed study of the intra-observation(s) QPE properties on short timescales (Miniutti et al. in preparation).

## V. QPE spectral evolution

We study the QPE spectral evolution during the highest-quality, longest observation (XMM4). Using phase-dependent good-time-intervals, we extract phase-resolved source and background spectra throughout the cycle. We divide the cycle in 7 phase-intervals with two spectra during the QPE rise ($R_1$ and $R_2$), one spectrum at the QPE peak (P), two spectra during the decay ($D_1$ and $D_2$), one pre-QPE spectrum ($Q_{pre}$), and one post-QPE spectrum ($Q_{post}$), as shown in Figure 3. A short phase-interval between $Q_{post}$ and $Q_{pre}$ is ignored.

We group all spectra to a minimum of 25 counts per energy bin, and we fit the 7 phase-resolved spectra simultaneously. As no hard tail is detected in any of the 7 spectra (likely because of low signal to noise), we consider a simple one-component accretion disc model (*diskbb*). Physically, this corresponds to assuming that the QPE spectral evolution can be described by global mass accretion rate variation throughout the disc. We reach a good description of the data with $\chi^2 = 468$ for 436 dof, and the most relevant best-fitting parameters are reported in Extended Data Table 3a. As the QPE progresses, the disc temperature smoothly rises from ~ 50 eV up to ~ 120 eV, and then decays back to ~ 50 eV in the $Q_{post}$ spectrum, i.e. GSN 069 oscillates between a cold and a warm phase during QPEs, as expected based on the QPE energy-dependence (e.g. Extended Data Figure 5b). Applying the same model to the Chandra quiescent and peak spectra shows that the oscillation is between a ~ 80 eV and a ~ 120 eV phase. The lower QPE amplitude during the Chandra observation is then explained by a less abrupt spectral difference between the two phases.

Although the one-component *diskbb* spectral model provides a statistically fair description of the QPE spectral evolution, it appears to be physically implausible. This is because the highest 0.2-2 keV luminosity during QPEs is only a factor of ~ 2.4 higher than during the quiescent phase, despite a factor of ~ 2 increase in temperature, i.e. the expected L ~ $T^4$ relation is not satisfied. Hence, global mass accretion rate variability in a constant-area emitting disc fails to provide a physically consistent explanation of the QPE phenomenon.

We then consider a second spectral decomposition in which we assume the presence of two thermal models. The first component is again modelled with *diskbb*, but forced to be constant in both temperature and normalization throughout the cycle, and representing the constant emission from a stable (likely outer) region of the disc. The second component is allowed to vary during the cycle and represents emission from the variable (likely inner) accretion flow. From a physical point of view, the variable component may be associated with Comptonization in a warm, optically thick corona or with local (rather than global) mass accretion rate variations in different disc regions during the QPE cycle (e.g. different annuli that lighten up at different times, as possibly suggested by Figure 2b). Since in both cases the X-ray spectral shape is blackbody-like, for the sake of generality we model the variable component with a simple blackbody (*bbody* in XSPEC) without making strong assumptions on its physical origin. We obtain $\chi^2 = 457$ for 435 dof, with an improvement that is significant at the 99.9 per cent level (F-test) with respect to the one-component model discussed earlier (see Extended Data Table 3b). The constant disc is characterized by kT ~ 48 eV, while the variable blackbody temperature varies between ~ 50 eV and ~ 105 eV as shown in Extended Data Figure 6a. The X-ray luminosity evolution of the variable component is displayed in Extended data Figure 6b. The global 0.2-2 keV luminosity reaches a maximum of ~ $5 \times 10^{42}$ erg s$^{-1}$ during QPEs. The best-fitting SED evolution throughout the cycle is shown in Extended data Figure 6c (for the QPE rise) and 6d (QPE decay).

We note that the temperature of the warm phase is similar to that of the typical soft X-ray excess observed in the X-ray spectra of most radiatively-efficient AGN. This may indicate that QPEs are recurrent, short-lived oscillations between a one-temperature disc-dominated and a two-temperature disc-soft-excess state. In this context, the Chandra soft-excess-dominated quiescent spectrum may be representative of an intermediate stage of the spectral transition between the two states. This suggests that we are possibly witnessing, for the first time, the QPE-driven formation of the soft X-ray excess in an AGN in real time. If the overall luminosity of GSN 069 continues to drop, the presence of a stable ~100-200 eV soft excess and weak or no QPEs in future X-ray observations would demonstrate that GSN 069 has completed a QPE-driven transition from a purely disc-dominated to a typical soft excess AGN spectral state (besides the typical power law component). On the other hand, if the Chandra observation represents a sporadic excursion into a low flux level and the X-ray flux recovers, we may expect a disc-dominated spectrum associated with strong QPEs again in the future.

## VI. The X-ray/radio campaign

Following the X-ray QPEs detection in XMM3 and XMM4, we organised a DDT-based X-ray/radio campaign with Chandra, the ATCA, the VLA, and the MeerKAT radio facilities (Extended Data Table 1). The MeerKAT1 and ATCA2 exposures comprise one Chandra X-ray QPE each, while the VLA observation was performed during the Chandra quiescent phase. The ATCA1 and MeerKAT2 exposures are not strictly simultaneous with the Chandra observation, but the propagation of the detected QPE times predicts that no QPEs were present during the exposures.

GSN 069 is detected by the VLA with a flux density of 47 ± 8 µJy at a mean frequency of 6 GHz, by MeerKAT with 147 ± 7 µJy (MeerKAT1) and 156 ± 14 µJy (MeerKAT2) at 1.3 GHz, and by ATCA with 129 ± 17 µJy (ATCA1) and 120 ± 10 µJy (ATCA2) at 9 GHz. At 5.5 GHz, the source is not detected by ATCA down to a 95 % upper limit of 126 µJy/beam during the longer ATCA2 exposure, consistent with the VLA measurement (~ 50 µJy). However, the ATCA measurements should be taken with caution as ATCA is unable to resolve GSN 069 from a contaminating source seen in the VLA map 50 arcseconds away to the North-West and for which the VLA measures a flux density of 71 ± 10 µJy at 6 GHz. GSN 069 was also observed on 2017 November 2 with the VLA for a total on-source time of just under 5 minutes. A point source fit gives a flux density of 61 ± 25 µJy, consistent with that measured in February 2019. To search for potential radio variability (e.g. associated with X-ray QPEs), each data set was divided into shorter time-intervals, but no significant variability was detected (see Extended Data Figure 7). Based on the MeerKAT and VLA flux densities, the radio spectral index is ~ -0.7, suggesting optically thin synchrotron emission. This also confirms that the ATCA measurements are indeed heavily contaminated. The lack of detected radio variability during QPEs (though based on MeerKAT1 data only, see Extended Data Figure 7) suggests that QPEs are not associated with ejections, although we cannot exclude that radio variability is present but undetected because the evolving radio components overlap with each other on the probed timescales.

The 6 GHz flux density during the 2019 VLA observation translates into a radio luminosity of $L_{radio}$ ~ $1.9 \times 10^{36}$ erg s$^{-1}$ (corresponding to an estimated jet power[52] $P_{jet}$ ~ $1.2 \times 10^{41}$ erg s$^{-1}$). $L_{radio}$ can be combined with the observed 2-10 keV X-ray luminosity during the simultaneous Chandra observation to obtain an estimate of the BH mass in GSN 069 using the fundamental plane of BH accretion[53]. We infer a BH mass of $M_{BH}$ ~ $2 \times 10^6$ $M_\odot$, i.e. about 5 times higher than that estimated from X-ray spectral analysis. We must point out, however, that the GSN 069 SED is remarkably different from that of the typical AGN used to build the fundamental plane. In particular, GSN 069 is extremely X-ray weak above 2 keV with respect to the typical AGN population, i.e. the BH mass obtained from the fundamental plane is likely over-estimated.

## VII. Possible interpretations of the QPE phenomenon

The observed oscillations between cold and warm phases are highly reminiscent of limit-cycle oscillations induced by instabilities of the accretion flow[19,20,54,55,56,57]. The QPE energy-dependence (see Figure 2 and Extended Data Figure 5) is also consistent with expectations from a limit-cycle instability in which a heating/cooling front propagates in the inner accretion flow. As all accretion timescales depend on the orbital one, the evolution is faster in the innermost, hotter region than in outer colder ones. This overall description appears to be consistent with e.g. Figure 2b, where we demonstrate that QPEs measured at higher X-ray energies (i.e. higher temperatures) peak earlier and evolve faster than QPEs measured at lower energies.

Let us then briefly consider an instability-driven interpretation of the observed QPE phenomenon. We refer here to the classical radiation pressure instability, but we point out that other types of instabilities could be relevant (e.g. related to magnetic fields and/or including disc-corona mass/energy exchange if QPEs are associated with Comptonization). Considering for simplicity the XMM4 observation, the average QPE properties, as obtained from the full 0.2-2 keV band light curve, are:

$T_{dur}$ = FWHM ~ 2,050 s;

$T_{rise}$ ~ $T_{decay}$ = 2 σ ~ 1,740 s;

$T_{rec}$ ~ 32,150 s,

so that the time spent in the cold phase is $T_{cold} = T_{rec} - T_{dur}$ = 30,100 s.

Our definition of $T_{rise}$ ~ $T_{decay}$ is somewhat arbitrary, but has very little impact on the following discussion. The physical accretion flow timescales - namely the dynamical (orbital), thermal, and viscous diffusion timescales - are

$t_{dyn}$ ~ $(GM_{BH}/R^3)^{-1/2}$;

$t_{th}$ ~ $\alpha^{-1} t_{dyn}$;

$t_{visc}$ ~ $\alpha^{-1} (H/R)^{-2} t_{dyn}$ = $(H/R)^{-2} t_{th}$,

where R is the radius, $M_{BH}$ is the BH mass, α ≤ 1 is the viscosity parameter, and H/R is the accretion flow scale-height. Finally, the timescale for the propagation of perturbations in the flow is $t_{prop}$ ~ $\alpha^{-1} (H/R)^{-1} t_{dyn}$.

As an order-of-magnitude estimate, $T_{cold}$ and $T_{dur}$ are naturally associated with the viscous diffusion timescale in the cold and warm phases respectively [54,55,56,57]. Assuming that the viscosity parameter α is the same in the two phases[56], and evaluating all accretion flow timescales at the same radius (the outer edge of the unstable region) yields $(H/R)_{warm} \sim (T_{cold} T_{dur}^{-1})^{1/2} (H/R)_{cold}$, so that $(H/R)_{warm} \sim 3.8 (H/R)_{cold}$, consistent with the theoretical expectation of an inflated accretion flow in the unstable region. On the other hand, the QPE rise (decay) cannot be faster than the thermal timescale, and may be instead related to the propagation of the perturbation front in the cold (warm) phase[57]. However, in the latter case, as H/R is different in the cold and warm phases, one would expect $T_{rise} \neq T_{decay}$. The very symmetric nature of the QPE profile thus suggest to identify $T_{rise} \sim T_{decay}$ with the thermal timescale (independent of H/R). Therefore we infer $(H/R)_{cold} \sim (T_{rise} T_{cold}^{-1})^{1/2} \sim 0.24$, and $(H/R)_{warm} \sim 0.9$.

Detailed numerical simulation of the radiation-pressure instability[56] show that $M_{BH}$ can be written in terms of the ratio A between the maximum and minimum bolometric luminosities, recurrence time $T_{rec}$, and viscosity parameter α as

$M_{BH} = 0.45 \, T_{rec}^{0.87} \, A^{-0.72} \, (\alpha/0.02)^{1.88}$,

where $M_{BH}$ is in units of $M_\odot$, and $T_{rec}$ in seconds. Using the GSN 069 parameters, we have that $M_{BH} \sim 3.1 \times 10^3 \, (\alpha/0.02)^{1.88} \, M_\odot$ which, assuming $M_{BH} \sim 4 \times 10^5 \, M_\odot$, yields $\alpha \sim 0.26$ in GSN 069. The QPE energy-dependence and, in particular, the lack of QPEs below $\sim 0.1$ keV suggest that the instability is confined within the innermost regions of the accretion flow (see Extended Data Figure 5a). Let us call $R_{unst}$ the outer edge of the unstable region. After correcting the observed recurrence time for both systemic and gravitational redshift (at $R_{unst}$), we can use the equivalence between $T_{cold}$ and $t_{visc}$ in the cold phase to derive an estimate of $R_{unst}$. Assuming $M_{BH} \sim 4 \times 10^5 \, M_\odot$, $(H/R)_{cold} \sim 0.24$, and $\alpha \sim 0.26$ as derived above, and evaluating $t_{visc}$ at $R_{unst}$, yields $R_{unst} \sim 36 \, R_g$.

We conclude that, if QPEs are instability-driven: $H/R \sim 0.24$ in the quiescent/cold phase, $H/R \sim 0.9$ in the warm phase at the QPE peak, $\alpha \sim 0.26$ (assumed to be the same in both phases), and the unstable region is confined within the innermost 30-40 $R_g$ of the accretion flow. Obviously, as all estimates are based on the comparison between observed time intervals and theoretical timescales, they have to be considered as simple and rough order-of-magnitude estimates.

As mentioned, the BH X-ray binaries GRS 1915+105 and IGR J17091-3624 in their heartbeat states[16,17,18] exhibit similar variability properties as GSN 069. The recurrence times for the heartbeat of GRS 1915+105 is highly variable and, typically, of the order of 10-1000 s. IGR J17091-3624 heartbeats have shorter recurrence time, typically by one order of magnitude with respect to GRS 1915+105. Although the BH masses in GRS 1915+105 and IGR J17091-3624 are possibly different, as we are here interested in order-of-magnitude estimates, we assume that they both have $M_{BH} \sim 10 \, M_\odot$. For the fiducial BH mass of GSN 069, $T_{rec}^{(GSN\ 069)}$ translates into a predicted recurrence times of the order of $\sim$ 1 s in the X-ray binary case. Given that the BH mass in GSN 069 is likely associated with a factor of a few uncertainty and that α, H/R, and $R_{unst}$ are unlikely to be exactly the same in the three systems, it does not seem implausible that the recurrence times can be made consistent with each other, i.e. that we may be observing the same or similar type of instability in all systems[21,22,57].

**(Some) alternative interpretations**

An instability-driven interpretation of the QPE phenomenon appears plausible given the observed QPEs timing and spectral properties. However, this is a new phenomenon, and it is clear that alternative interpretations should be explored. Here we simply mention a few potentially interesting scenarios, being aware that our discussion is far from exhaustive.

When quasi-periodic phenomena are observed in accreting systems, orbital motion is an obvious possibility. An orbiting structure with rest-frame blackbody-like spectrum may reproduce the observed QPE spectral evolution via Doppler effects. However, assuming the fiducial $M_{BH} = 4 \times 10^5 \, M_\odot$ and identifying the (rest-frame) recurrence time with the orbital period sets a scale of ~ 190 $R_g$. At such large distances, orbital motion would imprint a quasi-sinusoidal low-amplitude modulation rather than the observed abrupt, short-lived QPEs. We conclude that an orbiting-spot-like model is an unlikely explanation for the observed QPEs.

Another possibility is precession of an inner torus-like accretion flow. The lack of radio variability during X-ray QPEs (see §VI) makes it unlikely that variability is induced by jet precession. One may instead imagine a geometry in which the quiescent ~ 50 eV emission originates in the outer accretion disc, while the innermost regions comprise a geometrically thick torus-like flow with an inner X-ray-emitting funnel. A misaligned disc-torus geometry around a rotating Kerr BH is subject to Lense-Thirring precession of the inner torus. Torus precession could then produce high-amplitude variability as the funnel emission recurrently moves in and out of the line of sight. The Lense-Thirring period $T_{LT}$ depends on BH mass and spin, and on the torus size (in particular its outer edge $R_{tor}$). $T_{LT} \sim T_{rec} \sim 30$ ks can be obtained for reasonable values of $R_{tor} \sim$ 10-30 $R_g$ for a broad range of BH spin values[58] which makes this scenario worth exploring in the future. We note that the small observed QPE duty cycle (Δ ~ 6%) implies a very narrow X-ray-emitting funnel.

High amplitude recurrent variability events may also be induced by the interaction between an existing accretion disc and an orbiting secondary body. A similar scenario can successfully explain the optical variability of OJ 287[59]. In that system, optical outbursts occur every ~ 12 years with two peaks per cycle separated by 1-2 years, and the leading interpretation is that of a super-massive BH binary system in which the secondary passes through the accretion disk of the primary producing two optical outbursts per period[60]. In the case of GSN 069, the quasi-periodic nature of the X-ray variability implies an orbital configuration producing only one impact per orbit or a nearly circular orbit of the secondary. Whether this type of model can account for the QPE spectral evolution remains to be addressed in future work. We point out that no QPEs are observed during the XMM2 observation in 2014 December, despite the observation being long enough (~ 83 ks) to potentially detect two/three QPEs. This observational fact needs to be accounted for in any attempt of explaining the QPE phenomenon. In the context of disc-secondary body interactions, an initially compact disc (e.g. produced by a TDE) spreading out viscously may be large enough to intercept the secondary orbit only at late times, which may explain the lack of QPEs in the 2014 December XMM2 observation.

Another plausible scenario could be invoked in the context of a TDE-driven outburst for GSN 069. If the star was only partially disrupted, the surviving body could still orbit the BH and may produce QPEs either by a mechanism similar to that outlined above, or by Roche lobe overflow at closest approach, giving rise to enhanced quasi-periodic accretion events. However, the lack of QPEs in XMM2 and the consistency between orbital/dynamical parameters and observed properties/timescales needs to be addressed.

Exploring in detail these (and possibly other) explanations of the QPE phenomenon is beyond the scope of this Letter, and will be the subject of future dedicated work.

## VIII. QPEs in the wider context

In an attempt to identify other potential QPE candidates, it is obviously worthwhile to look for sources sharing the largest possible number of properties with GSN 069. As already noted in our previous work[4], 2XMM J123103.2+110648[14,61] (J1231) is one such source. It is another low BH mass Seyfert 2 galaxy that is characterized by an ultra-soft, disc-like X-ray spectrum with only weak power law emission and high amplitude, short timescale soft X-ray variability. Its long-term evolution is also similar to that of GSN 069 to the extent that it has been proposed as an example of long-lived TDE[62], as may be the case for GSN 069. Remarkably, a ~ 3.8 hr QPO has been detected in J1231[12,14]. It is worth mentioning that, in the context of a disc-instability interpretation of QPEs, limit-cycle oscillations are predicted to become low-amplitude quasi-periodic flickering[54,55] as the mass accretion rate drops, due to the shrinking size of the unstable region. The observed QPO in J1231 may then be the signature of a weak QPE, or of a QPE that does not fully develop. This also suggests the possibility that QPEs in GSN 069 may develop into a gentler lower-amplitude modulation (a QPO) in the future, as the mass accretion rate continues to drop.

Another similar source is RX J1301.9+2747[15,63] (J1301), a low BH mass AGN whose X-ray spectrum is dominated by a thermal disc with weak power law emission emerging above ~ 1 keV. The optical spectrum shows no signs of broad optical emission lines, as in GSN 069 and J1231. The most remarkable property of J1301 is the presence of short, high-amplitude X-ray flares that are observed in 2000 and 2009 observations with XMM-Newton and Chandra respectively[15]. The source is active at similar flux levels since the ROSAT era, and ROSAT data also indicate high amplitude and short timescale soft X-ray variability[64]. We have recently obtained an X-ray/radio observation to search for QPEs in J1301, and results will be reported elsewhere (Giustini et al. in preparation).

As mentioned, both GSN 069 and J1231 exhibit long-term X-ray decays that could be understood in terms of long-lived TDEs, perhaps associated with the disruption of evolved stars off the main sequence. This possibility may be viable also for J1301, as outbursts produced by the disruption of giant stars may last hundreds of years[39]. In fact, TDEs are generally characterized, at least transiently, by properties that are similar to those of GSN 069 at the time of first QPE detection. Thermal TDEs are disc-dominated with generally weak hard X-ray emission, and they are likely to cross the $\sim 0.5$ Eddington ratio regime at some point during their long-term evolution. Moreover, the resulting disc is likely too small to sustain a mature broad line region, which may explain why no optical broad lines are detected in GSN 069 (nor in J1231 and J1301). Long-lived TDEs may then be among the most promising candidates where to search for QPEs. More generally, the most secure QPE candidates should be looked for within a population of relatively high Eddington ratio, low BH mass AGN with soft X-ray emission characterized by a temperature significantly lower than that of the standard soft excess, weak power law emission above 1-2 keV, and no/weak optical broad emission lines.

On the other hand, the typical AGN population is characterized by significantly higher BH masses of the order of $10^7$-$10^9$ $M_\odot$. In massive AGN, the thermal disc does not extend up to soft X-rays and dies off in the extreme/far UV instead. Hence, prior to the appearance of QPEs, a GSN 069-like more massive AGN appears in the X-ray band as an extremely X-ray weak source with power-law-like X-ray spectrum and no soft X-ray excess. Being more luminous than its low BH mass counterpart, the optical-to-UV SED of such an AGN may reveal the nuclear continuum dominating above galactic emission. As mentioned, QPEs are best described by short timescale oscillations between a relatively cold disc and a warmer phase. The detectability of QPEs at X-ray energies in systems with higher BH mass than GSN 069 depends on whether the warm phase temperature is universal, i.e. related to the AGN soft X-ray excess, or rather associated with a given increase w.r.t. the quiescent level temperature (i.e. a factor of $\sim 2$ in GSN 069). In the former case, soft X-ray QPEs could be potentially detected in massive AGNs as recurrent excursions in soft-excess-dominated states, while in the latter case QPEs would likely be confined to the FUV/EUV band and hence more difficult to observe, although the higher luminosity and therefore dominance w.r.t. stellar processes in the UV may help. Multi-epoch X-ray observations of variable AGN with appearing/disappearing soft X-ray excess may be a key observational signature of QPEs in massive BH systems, and could confirm the association between the QPE warm phase and the AGN soft X-ray excess.

The extremely weak power law emission and lack of soft X-ray excess prior to QPEs detection that characterizes GSN 069 (as well as J1231 and J1301) may be a fundamental property for QPE to develop, especially if QPEs are instability-driven: numerical simulations of the radiation-pressure instability show that the X-ray coronae (responsible for the power law emission and, possibly, for the soft X-ray excess as Comptonization in a warm optically thick medium[65]) have a stabilizing effect, significantly reducing the instability amplitude, or even suppressing it altogether[66]. Note that outflows have generally a similar stabilising effect. Potential systems where to look for QPEs include therefore genuine (i.e. not absorbed) X-ray weak AGN with no apparent soft X-ray excess. If the lack of broad optical lines is crucial for QPEs, weak emission line (X-ray weak) AGN are also potentially interesting targets.

**QPEs and extreme variability AGNs**

As discussed in §VII, the observed QPE timescales may be used to derive the quiescent accretion flow scale-height as $H/R \sim 0.2$ in GSN 069, under the assumption that QPEs are due to limit-cycle instabilities. If the long-term accretion event in GSN 069 is TDE-driven, a mildly thick accretion flow is perhaps not surprising, as TDEs accretion flows are expected to be significantly thicker than AGN discs. This is in fact likely true for high Eddington ratio accretion flows more generally[67,68]. A thicker-than-standard accretion flow may be associated with radiation pressure dominance and/or magnetically elevated accretion[23,24]. However, in the radiation-pressure scenario, $H/R$ is a function of mass accretion rate, or Eddington ratio. QPEs are observed with similar recurrence time as in XMM3/XMM4 also during the Chandra observation, which is characterized by a lower Eddington ratio (see Extended Data Table 2 and the associated discussion in §III). As $T_{rec} \propto (H/R)^{-2}$, this suggests that $H/R$ is not a strong function of the Eddington ratio (or mass accretion rate), so that a magnetically supported disc[24] may be a more appropriate explanation for the suggested relatively high $H/R$.

We note that if the relatively thick disc we measure in GSN 069 is not source-specific but a more general result, more massive AGN can vary on shorter timescales than expected from the razor-thin standard disc assumption. As a simple and admittedly over-simplified exercise, the viscous diffusion timescale at $R \sim 200\ R_g$ for a typical $10^8\ M_\odot$ AGN with $H/R \sim 0.2$ and $\alpha \sim 0.2$ (as inferred in GSN 069, see §VII), is as short as $\sim$ few years not only in the X-ray, but even in the UV and optical (as we have evaluated the viscous timescale at $200\ R_g$). This may then be a plausible explanation for the rapid variability seen in a growing number of AGNs, including the so-called changing look AGNs[9,69,70,71].

Some changing-look AGNs (or other extreme variability objects) may even be explained naturally if they were actually experiencing QPEs, in which case the observed QPE properties would imply similar numbers of changing-look AGNs in the rise and decay phases, and an appearing/disappearing soft X-ray excess during the rise/decay phase.

The latter prediction is consistent with the long-term variability of the two changing-look AGNs Mrk 590[25,26,27,30] and Mrk 1018[28,29]. Over the past ~ two decades, the continuum emission from Mrk 590 has faded away (together with the optical, broad emission lines), and this transition is accompanied by the disappearance of the soft X-ray excess that characterized the X-ray spectrum in the high state[26]. Mrk 590 is recently awakening, and its X-ray spectrum shows that a weak soft excess has re-emerged[27], together with its optical broad emission lines[30]. On the other hand, Mrk 1018 changed from optical type 1.9 to 1 around 1980[28], and again from type 1 to 1.9 around 2010[29] with associated high amplitude optical to X-ray variability. The latest decline (~ 8 years duration) is relatively well monitored in the UV and X-rays by Swift and XMM-Newton, revealing that the decay and spectral type transition are accompanied by the disappearance of the soft X-ray excess[23]. Long-term monitoring of these and other potential QPE candidates may reveal recurrent QPEs, and could confirm the association between the QPE warm phase and the typical soft X-ray excess in AGNs, providing important clues on the origin and formation of this puzzling X-ray spectral component.

**References**

**31.** Saxton, R. D. et al., The first XMM-Newton slew survey catalogue: XMMSL1, *Astron. and Astrophys.*, **480**, 611-622 (2008)

**32.** Sault, R. J., Teuben, P. J., Wright, M. C. H., A Retrospective View of MIRIAD, *Astronom. Data Analysis Software. and Systems IV, ASP Conf. Series*, **77**, 433-436 (1995)

**33.** McMullin, J. P., Waters, B., Shiebel, D., Young, W., Golap, K., CASA architecture and applications, *Astronom. Data Analysis Software. and Systems XVI, ASP Conf. Series*, **376**, 127-130 (2007)

**34.** Williams, P. K. G., Clavel, M., Newton, E. and Ryzhkov, D., pwkit: astronomical utilities in Python, *Astrophys. Source Code Library*, record ascl:1704.001 (2017)

**35.** Offringa, A. et al., WSCLEAN: an implementation of a fast, generic wide-field imager for radio astronomy, *Mon. Not. R. Astron. Soc.*, **444**, 606619 (2014)

36. Kenyon, J. S., Smirnov, O. M., Grobler, T. L., & Perkins, S. J., CUBICAL - fast radio interferometric calibration suite exploiting complex optimization, *Mon. Not. R. Astron. Soc.*, **478**, 2399-2415 (2018)

37. Tasse, C. et al., Faceting for direction-dependent spectral deconvolution, *Astron. and Astrophys.*, **611**, A87 (2018)

38. Farrell, S. A., Webb, N. A., Barret, D., Godet, O. and Rodrigues J. M., An intermediate-mass black hole of over 500 solar masses in the Galaxy ESO 243-49, *Nature*, **460**, 73-75 (2009)

39. MacLeod, M., Guilochon, J. and Ramirez-Ruiz, E., The Tidal Disruption of Giant Stars and their Contribution to the Flaring Supermassive Black Hole Population, *Astrophys. J.*, **757**, 134 (2012)

40. Pickles, A. J., A stellar spectral flux library: 1150-25000 Å, *Publ. of the Astron. Soc. Of Japan*, **110**, 863-878 (1998)

41. Mitsuda, K. et al., Energy spectra of low-mass binary X-ray sources observed from TENMA, *Publ. of the Astron. Soc. of Japan*, **36**, 741-759 (1984)

42. Wilms, J., Allen, A. and McCray, R., On the Absorption of X-Rays in the Interstellar Medium, *Astrophys. J.*, **542**, 914-924 (2000)

43. Piconcelli, E., The XMM-Newton view of PG quasars. I. X-ray continuum and absorption, *Astron. and Astrophys.*, **432**, 15-30 (2005)

44. Reeves, J. et al., On why the iron K-shell absorption in AGN is not a signature of the local warm/hot intergalactic medium, *Mon. Not. R. Astron. Soc.*, **385**, L108-L112 (2008)

45. Kalberla, P. M. W., The Leiden/Argentine/Bonn (LAB) survey of Galactic HI. Final data release of the combined LDS and IAR surveys with improved stray-radiation corrections, *Astron. and Astrophys.*, **440**, 775-782 (2005)

46. Rees, M. J., Tidal disruption of stars by black holes of 10 to the 6$^{th}$ -10 to the 8$^{th}$ solar masses in nearby galaxies, *Nature*, **333**, 523-528 (1988)

47. Phinney, E. S., Manifestations of a massive black hole in the galactic center, *Proc. Of the 136$^{th}$ Symp. of the IAU eited by M. Morris*, *Kluwer Acad. Publishers*, **136**, 543-552 (1989)

48. Yuan, W., Liu, B. F., Zhou, H. and Wang, T. G., X-ray observational signature of a black hole accretion disk in an active galactic nucleus RX J1633+4718, *Astrophys. J.*, **723**, 508-513 (2010)

49. Vasudevan, R. V. and Fabian, A. C., Simultaneous X-ray/optical/UV snapshots of active galactic nuclei from XMM-Newton: spectral energy distributions for the reverberation mapped sample, *Mon. Not. R. Astron. Soc.*, **392**, 1124-1140 (2009)

50. Ranalli, P., Comastri, A. and Setti, G., The 2-10 keV luminosity as a Star Formation Rate indicator, *Astron. and Astrophys.*, **399**, 39-50 (2003)

51. Kennicutt, R. C., Jr., Star Formation in Galaxies Along the Hubble Sequence, *Ann. Rev. of Astron. and Astrophys.*, **36**, 189-232 (1998)

52. Merloni, A. and Heinz, S., Measuring the kinetic power of active galactic nuclei in the radio mode, *Mon. Not. R. Astron. Soc.* **381**, 589-601 (2007)

53. Merloni, A., Heinz, S. and di Matteo T., A fundamental plane of black hole activity, *Mon. Not. R. Astron. Soc.* **345**, 1057-1076 (2003)

54. Merloni, A. and Nayakshin S., On the limit-cycle instability in magnetized accretion discs, *Mon. Not. R. Astron. Soc.*, **372**, 728-734 (2006)

55. Grzędzielski, M., Janiuk, A., Czerny, B. and Wu, Q., Modified viscosity in accretion disks. Application to Galactic black hole binaries, intermediate mass black holes, and active galactic nuclei. *Astron. Astrophys.*, **603**, A110 (2017)

56. Janiuk, A., Czerny, B., Siemiginowska, A. and Szczerba, R., On the turbulent α-disks and the intermittent activity in active galactic nuclei, *Astrophys. J.*, **602**, 595-602 (2004)

57. Nayakshin, S., Rappaport, S. and Melia, F., Time-dependent disk models for the microquasar GRS1915+105, *Astrophys. J.*, **535**, 798-814 (2000)

58. Ingram, A., Done, C. and Fragile, P. C., Low-frequency quasi-periodic oscillations spectra and Lense-Thirring precession, *Mon. Not. R. Astron. Soc.* **397**, L101-L105 (2009)

59. Sillanpää, A., Haarala, S., Valtonen, M.J., Sundelius, B. and Byrd, G. G., OJ 287: binary pair of supermassive black holes, *Astrophys. J.*, **325**, 628-634 (1998)

60. Valtonen, M. et al., A massive binary black-hole system in OJ287 and a test of general relativity, *Nature*, **452**, 851-853 (2008)

61. Ho, L. C., Kim, M. and Terashima, Y., The Low-mass, Highly Accreting Black Hole Associated with the Active Galactic Nucleus 2XMM J123103.2+110648, *Astrophys. J.*, **759**, L16 (2012)

62. Lin, D. et al., Large decay of X-ray flux in 2XMM J123103.2+110648: evidence for a tidal disruption event, *Mon. Not. R. Astron. Soc.* **468**, 783-789 (2017)

63. Shu, X.W. et al., Central engine and host galaxy of RXJ 1301.9+2747: a multiwavelength view of a low-mass black hole active galactic nuclei with ultra-soft X-ray emission, *Astrophys. J.*, **837**, 3 (2017)

64. Dewangan, G. C., Singh, K. P., Mayya, Y. D. and Anupama, G. C., Active nucleus in a post-starburst galaxy: KUG 1259128, *Mon. Not. R. Astron. Soc.* **318**, 309-320 (2000)

65. Petrucci, P.-O., Testing warm Comptonization models for the origin of the soft X-ray excess in AGN, *Astron. And Astrophys.*, **611**, A59 (2018)

66. Janiuk, A., Czerny, B. and Siemiginowska, A., Radiation pressure instability driven variability in the accreting black holes, *Astrophys. J.*, **576**, 908-922 (2002)

67. Sądowski, A. and Narayan, R., *Powerful radiative jets in supercritical accretion discs around non-spinning black holes, Mon. Not. R. Astron. Soc.* **453**, 3213-3221 (2015)

68. Jiang, Y.-F., Stone, J. M. and Davis, S. W., *A global three-dimensional radiation magneto-hydrodynamic simulation of super-Eddington accretion disks, Astrophys. J.* **796**, 106 (2014)

69. Matt, G., Guainazzi, M. and Maiolino R., Changing look: from Compton-thick to Compton-thin, or the rebirth of fossil active galactic nuclei, *Mon. Not. R. Astron. Soc.* **342**, 422-426 (2003)

70. Shappee, B. J., The man behind the curtain: X-rays drive the UV through NIR variability in the 2013 active galactic nucleus outburst in NGC 2617, *Astrophys. J.* **788**, 48 (2014)

71. LaMassa, S. et al., The discovery of the first "changing-look" quasar: new insights into the physics and phenomenology of active galactic nucleus, *Astrophys. J.* **800**, 144 (2015)

**Extended Data Table 1 | Summary of the observations used in this work**

| Observation | ObsID | Start date | Exposure | Observation | ObsID | Start date | Exposure |
|---|---|---|---|---|---|---|---|
| **XMM1** | 0657820101 | 2010-12-02 | 11 ks | **ATCA1** | CX425 | 2019-01-26 | 19 ks |
| **XMM2** | 0740960101 | 2014-12-05 | 83 ks | **MeerKAT1** | 1550159669 | 2019-02-14 | 11 ks |
| **HST/STIS1** | 13815 | 2014-12-14 | 8 ks | **Chandra** | 22096 | 2019-02-14 | 73 ks |
| **XMM3** | 0823680101 | 2018-12-24 | 45 ks | **VLA** | 19A-454 | 2019-02-14 | 13 ks |
| **HST/STIS2** | 15442 | 2018-12-31 | 8 ks | **ATCA2** | CX425 | 2019-02-14 | 38 ks |
| **XMM4** | 0831790701 | 2019-01-16 | 133 ks | **MeerKAT2** | 1550233871 | 2019-02-15 | 11 ks |

We also use data from a short archival VLA observation performed on 2017 November 2 (17B-027; PI: X.W. Shu), see §I and §VI for details.

**Extended Data Table 2 | Most relevant best-fitting parameters for the long-term evolution of GSN 069 using X-ray spectra from which time intervals containing QPEs are excluded**

| TBABS × ZXIPCF[*] × (DISKBB + POWERLAW) | | | | | | $\chi^2$/dof = 282/254 |
|---|---|---|---|---|---|---|
| Parameter | [Units] | XMM1 | XMM2 | XMM3 | XMM4 | Chandra |
| $N_H$ | [$10^{20}$ cm$^{-2}$] | 5.5 ± 1.1 | - | - | - | - |
| Γ | | 1.8[f] | - | - | - | - |
| kT | [eV] | 63 ± 3 | 53 ± 2 | 48 ± 3 | 47 ± 3 | 82 ± 16 |
| $F_{0.3-2}$ | [$10^{-12}$ erg s$^{-1}$ cm$^{-2}$] | 1.3 ± 0.1 | 0.69 ± 0.05 | 0.34 ± 0.03 | 0.34 ± 0.03 | 0.09 ± 0.03 |
| $F_{2-10}$ | [$10^{-14}$ erg s$^{-1}$ cm$^{-2}$] | 1.7 ± 0.6 | 0.9 ± 0.2 | 2.0 ± 0.6 | 1.9 ± 0.5 | 1.1 ± 0.3 |
| $L_{DISKBB}$ | [$10^{42}$ erg s$^{-1}$] | 11.0 ± 0.9 | 5.0 ± 0.4 | 3.0 ± 0.3 | 3.1 ± 0.3 | 0.15 ± 0.05 |
| $L_{PL}$ | [$10^{40}$ erg s$^{-1}$] | 1.2 ± 0.4 | 0.7 ± 0.2 | 1.4 ± 0.4 | 1.4 ± 0.4 | 0.8 ± 0.2 |
| $L_{Bol}$ | [$10^{43}$ erg s$^{-1}$] | ~ 4.8 | ~ 3.0 | ~ 2.3 | ~ 2.3 | ~ 0.09[**] |
| $L_{Bol}$ / $L_{Edd}$ | | ~ 0.95 | ~ 0.60 | ~ 0.46 | ~ 0.46 | ~ 0.02[**] |

[*] The ZXIPCF model is only applied to the XMM1 data, see caption for details.

[f] The photon index Γ is largely unconstrained by the data above 1-2 keV, so that it is fixed to Γ = 1.8 at all epochs.

[**] We point out that $L_{Bol}$ (and hence $L_{Bol}$ / $L_{Edd}$) at the Chandra epoch is significantly more uncertain than that derived from the other observations, as we are effectively fitting a soft excess rather than disc emission (see Methods §III, where we derive a range of $L_{Bol}$ ~ 0.09-1.1 × $10^{43}$ erg s$^{-1}$).

An intrinsic ionized absorber (the ZXIPCF model) is required by the XMM1 data. Its presence is not statistically significant in any of the other observations, and the model is therefore applied only to the XMM1 spectrum. The XMM1 absorber parameters are $N_H$ = (5.2 ± 0.36) × $10^{21}$ cm$^{-2}$ and log ξ = 0.35 ± 0.25. All fluxes are as observed, and luminosities are unabsorbed. $L_{DISKBB}$ and $L_{PL}$ are the 0.2-2 keV and 2-10 keV luminosities from the disc and power law models respectively. The bolometric luminosity $L_{Bol}$ is estimated from the extrapolation of the best-fitting models down to $10^{-5}$ keV (~ 124 μm). The Eddington ratio ($L_{Bol}$ / $L_{Edd}$) is estimated assuming a BH mass $M_{BH}$ = 4 × $10^5$ $M_\odot$ (see Methods §III). Errors represent the 90 % confidence level intervals as obtained from X-ray spectral fits.

**Extended Data Table 3 | Most relevant best-fitting parameters for the QPE phase-resolved spectral analysis during the XMM4 observation**

a

| TBABS × DISKBB | | | | | | | | χ²/dof = 468/436 |
|---|---|---|---|---|---|---|---|---|
| Parameter | [Units] | $Q_{pre}$ | $R_1$ | $R_2$ | P | $D_1$ | $D_2$ | $Q_{post}$ |
| $N_H$ | [$10^{20}$ cm$^{-2}$] | 3.5 ± 1.2 | - | - | - | - | - | - |
| kT | [eV] | 50 ± 2 | 82 ± 5 | 126 ± 5 | 122 ± 4 | 96 ± 3 | 67 ± 3 | 53 ± 3 |
| $F_{0.3-2}$ | [$10^{-12}$ erg s$^{-1}$ cm$^{-2}$] | 0.30 ± 0.01 | 0.47 ± 0.03 | 1.36 ± 0.05 | 2.31 ± 0.05 | 1.48 ± 0.05 | 0.60 ± 0.03 | 0.37 ± 0.03 |
| L / L$^{(Qpre)}$ | | 1 | ~ 0.7 | ~ 1.3 | ~ 2.4 | ~ 1.9 | ~ 1.2 | ~ 1.1 |

b

| TBABS × ( DISKBB + BBODY ) | | | | | | | | χ²/dof = 457/435 |
|---|---|---|---|---|---|---|---|---|
| Parameter | [Units] | $Q_{pre}$ | $R_1$ | $R_2$ | P | $D_1$ | $D_2$ | $Q_{post}$ |
| $N_H$ | [$10^{20}$ cm$^{-2}$] | 4.2 ± 1.1 | - | - | - | - | - | - |
| $kT_{const}$ | [eV] | 48 ± 2 | - | - | - | - | - | - |
| $kT_{var}$ | [eV] | 50$^f$ | 90 ± 6 | 105 ± 4 | 99 ± 4 | 82 ± 3 | 65 ± 4 | 55 ± 7 |
| $L_{const}$ | [$10^{42}$ erg s$^{-1}$] | 2.0 ± 0.2 | - | - | - | - | - | - |
| $L_{var}$ | [$10^{42}$ erg s$^{-1}$] | ≲ 0.2 | 0.36 ± 0.03 | 1.53 ± 0.09 | 3.0 ± 0.1 | 2.2 ± 0.1 | 0.75 ± 0.05 | 0.24 ± 0.02 |

Table **a** refers to the one-component variable DISKBB model applied to all phase-resolved spectra. In the last row, we report the ratio between the observed phase-dependent 0.2-2 keV luminosity and that during the pre-QPE phase ($Q_{pre}$). Together with the corresponding measured disc temperatures, this shows that the variable emission does not follow the L ~ T$^4$ relation, and thus that global mass accretion rate variability is not a viable explanation of the QPE phenomenon. table **b** reports results for a two-component thermal model where the DISKBB model parameters are forced to be phase-independent (i.e. constant throughout the cycle) representing a stable (likely outer) disc, and the QPE evolution is described by an additional variable BBODY component. $L_{const}$ and $L_{var}$ in table **b** refer to the 0.2-2 keV unabsorbed luminosities of the constant (DISKBB) and variable (BBODY) thermal models, characterized by $kT_{const}$ and $kT_{var}$ respectively. Errors represent the 90 per cent confidence level uncertainties.

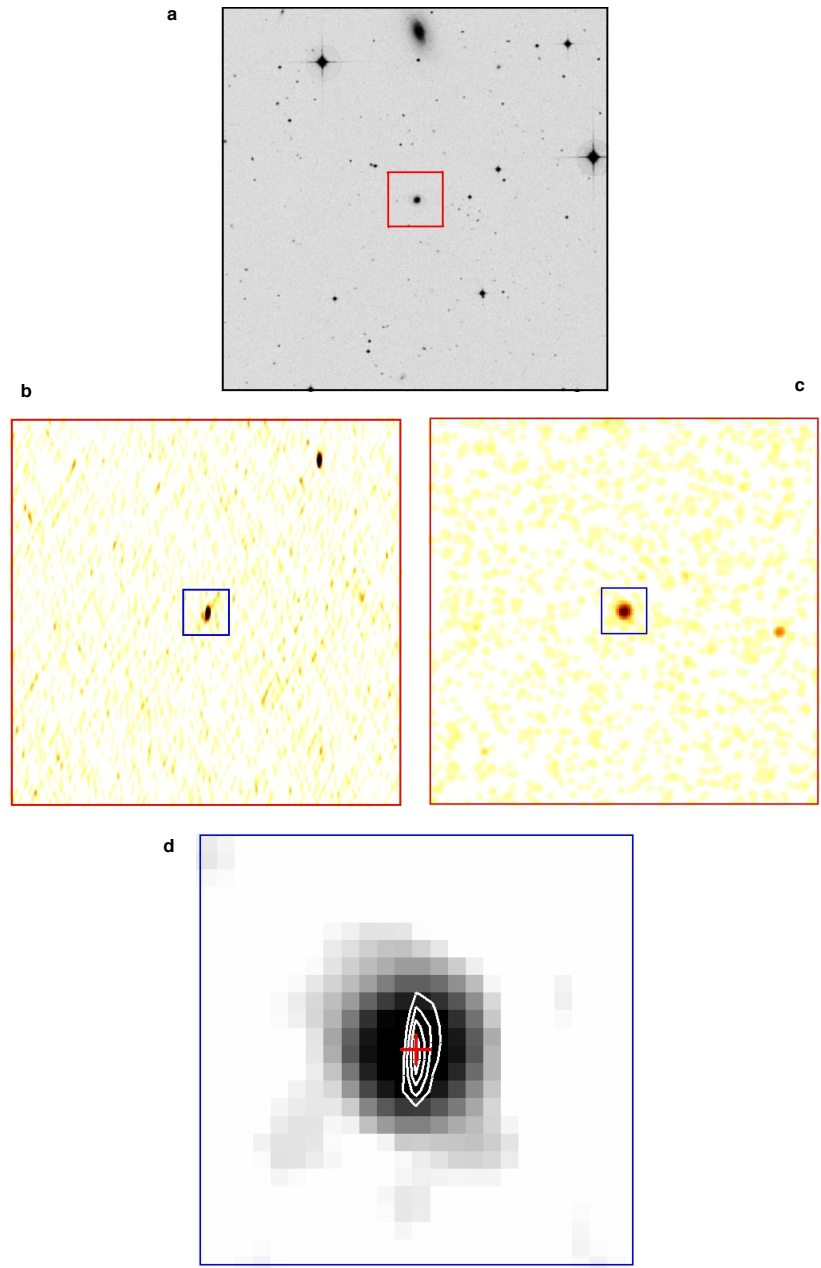

**Extended Data Figure 1 | Position of GSN 069 as inferred from optical, X-ray and radio images**
In **a** we show a 12′×12′ region from the Digitized Sky Survey (DSS) centered on the 2MASS position of GSN 069 (RA 01:19:08.663; DEC -34:11:30.52). The galaxy to the North of GSN 069 and close to the edge of the field is ESO 352-G41. The two bright stars in the field are CD-34 503 and CD-34 498. The red box size is 1.7′×1.7′. Panels **b** and **c** show the same 1.7′×1.7′ region as imaged by the VLA at 6 GHz and by Chandra in the 0.4-2 keV band respectively. The blue box size in **b** and **c** is 12″×12″ and panel **d** shows the Chandra X-ray image of that very same region, with superimposed VLA contours (white) as well as the 2MASS position (red cross) as reference. No boresight correction was applied to the Chandra data.

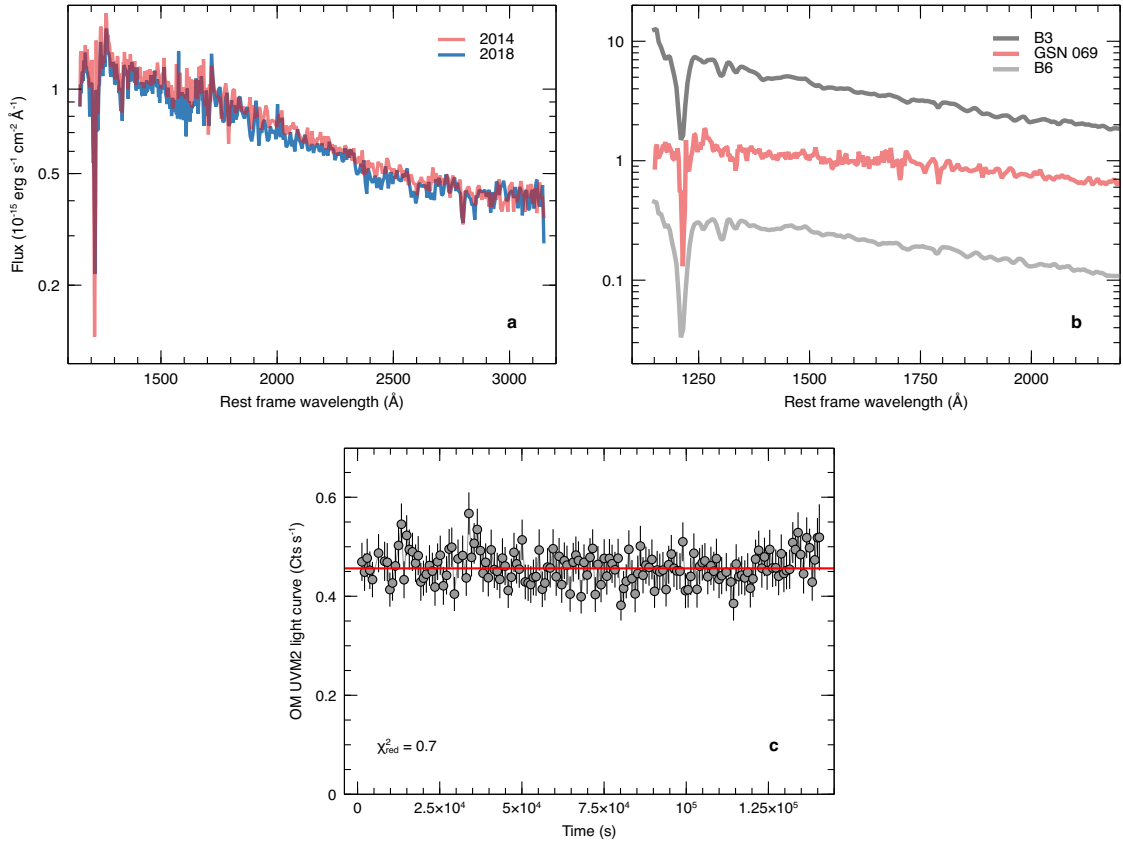

**Extended Data Figure 2 | UV spectrum and variability of GSN 069 from HST and OM observations**
In panel **a**, we show the 2014/2018 HST/STIS spectra that have been taken quasi-simultaneously with XMM2 (2014) and XMM3 (2018). In **b**, the 2014 STIS spectrum is compared with that of intermediate-type main sequence stars (B3 and B6) from the Pickles Atlas, demonstrating that the UV spectrum is strongly contaminated by starlight, and possibly dominated by a relatively young nuclear stellar cluster. Panel **c** shows the OM light curve in the UVM2 filter (~ 231 nm) during the XMM4 observation. The OM light curve shows no variability, with a reduced $\chi^2$ of 0.7 when fitted with a constant, despite the simultaneous X-ray QPEs (Figure 1b). The STIS spectra as well as the B3 and B6 spectra from the Pickles Atlas are displayed with no uncertainties, while errors in **c** represent the 1-σ confidence intervals.

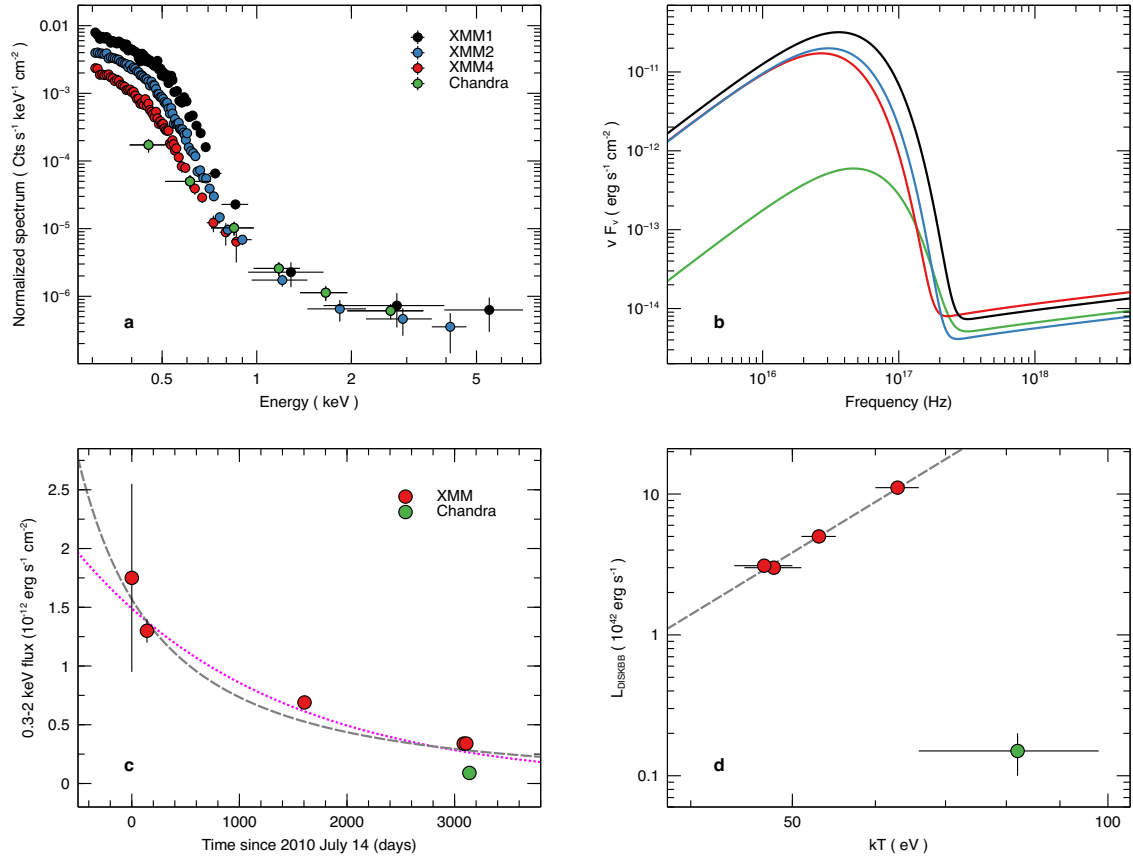

**Extended Data Figure 3 | The X-ray long-term evolution of GSN 069**

The X-ray spectra from the XMM1, XMM2, XMM4, and Chandra observations, excluding time-intervals containing QPEs, are shown in panel **a**. All spectra have been divided by the corresponding detector effective area to ease comparison. The XMM3 spectrum is not shown as it is basically superimposed on the XMM4 one. Spectra have been slightly re-binned for visual clarity. Panel **b** shows the best-fitting SEDs according to the best-fitting models presented in Extended Data Table 2. The 0.3-2 keV flux evolution of GSN 069 since first X-ray detection is shown in **c**, including the XMM-Newton slew data point. The dashed (grey) line is a power law decay model with index fixed at -5/3, while the dotted (magenta) line is an exponential decay law with best-fitting e-folding timescale of ~ 5 yr. In panel **d** we show the measured *diskbb* model 0.2-2 keV luminosity as a function of disc temperature (see Extended Data Table 2). The dashed line is the best-fitting relation $L \sim T^{4.5 \pm 0.5}$ to the XMM-Newton data only, consistent with constant-area blackbody emission ($L \sim T^4$). The Chandra data point (green) is far off the $L \sim T^4$ relation, its temperature being too hot to be ascribed to disc emission for the given luminosity. Errors in **a** represent the 1-σ confidence intervals, while error bars in **c** and **d** represent the 90 per cent confidence intervals as obtained from X-ray spectral fitting (Extended Data Table 2). Some of the error bars are smaller than the symbol size.

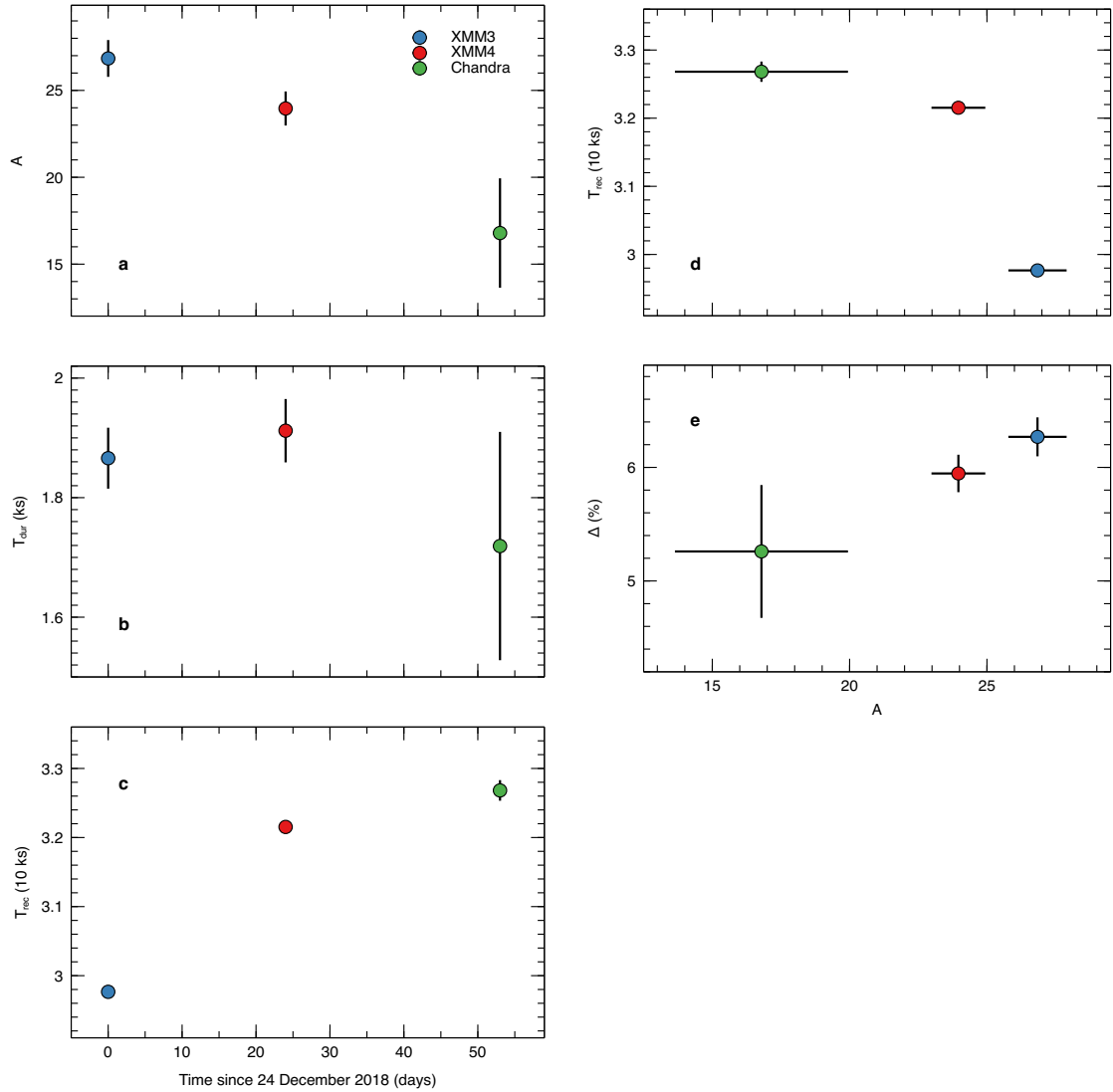

**Extended Data Figure 4 | Observation-averaged QPE long-term evolution**

In the left panels, we show the time-evolution of the QPE amplitude A (**a**), duration $T_{dur}$ (**b**), and recurrence time $T_{rec}$ (**c**) since first QPE detection in XMM3. All quantities are average over each X-ray observation. In the right panels, $T_{rec}$ (**d**) and the duty cycle $\Delta$ (**e**) are shown as a function of the QPE amplitude A. Errors bars represent the 1-$\sigma$ confidence intervals. Some error bars are smaller than the symbol size.

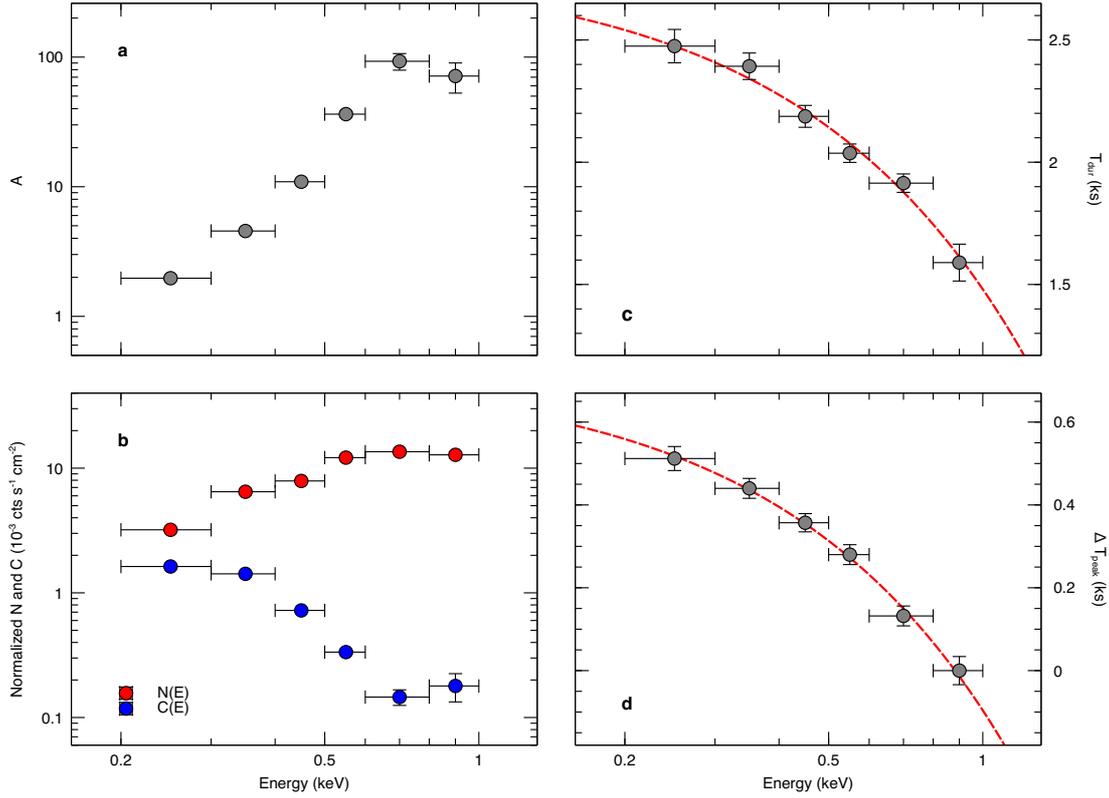

**Extended Data Figure 5 | QPE energy-dependence from the XMM4 observation**

In panel **a**, we show the QPE amplitude (A=N/C) as a function of energy. The maximum A = 93 ± 14 is reached in the 0.6-0.8 keV band. Panel **b** shows N=N(E) and C=C(E) normalized to the detector effective area in each energy bin, i.e. the QPE peak (N) and quiescent level (C) photon spectrum. The QPE duration $T_{dur}$ and QPE peak time delay $\Delta T_{peak}$ as a function of energy are shown in panels **c** and **d** respectively, together with the best-fitting linear relations (see §IV). The peak time delay is computed with respect to the full 0.2-2 keV light curve, and the resulting lags are shifted so that the 0.8-1 keV band has zero delay. Errors represent the 1-σ confidence intervals, some of the error bars being smaller than the symbol size.

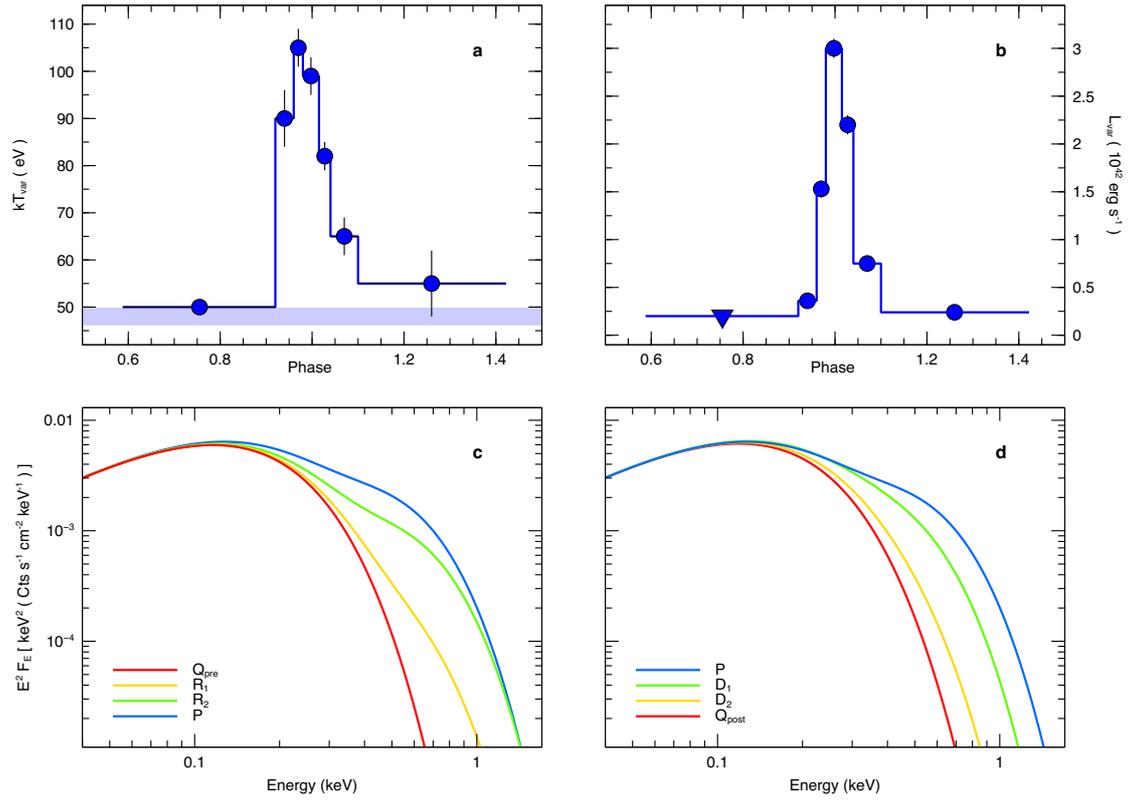

**Extended Data Figure 6 | QPE phase-resolved spectroscopy from the XMM4 observation**
In panels **a** and **b**, we show the temperature and 0.2-2 keV luminosity of the variable blackbody component throughout the QPE cycle (see Extended Data Table 3b). In **a**, the shaded area represents the constant temperature of the stable (likely outer) accretion disc. Panels **c** and **d** show the corresponding SED evolution: **c** shows the best-fitting model during the QPE rise from quiescence ($Q_{pre}$) to QPE peak (P), while **d** shows the QPE decay evolution from peak (P) back to quiescence ($Q_{post}$). The model predicts no variability below 0.1 keV, in line with the expectations based on Extended Data Figures 5a and 5b. In panels **a** and **b**, errors represent the 90 per cent confidence intervals as obtained from X-ray spectral fits to the phase-resolved spectra (some of the error bars being smaller than the symbol size).

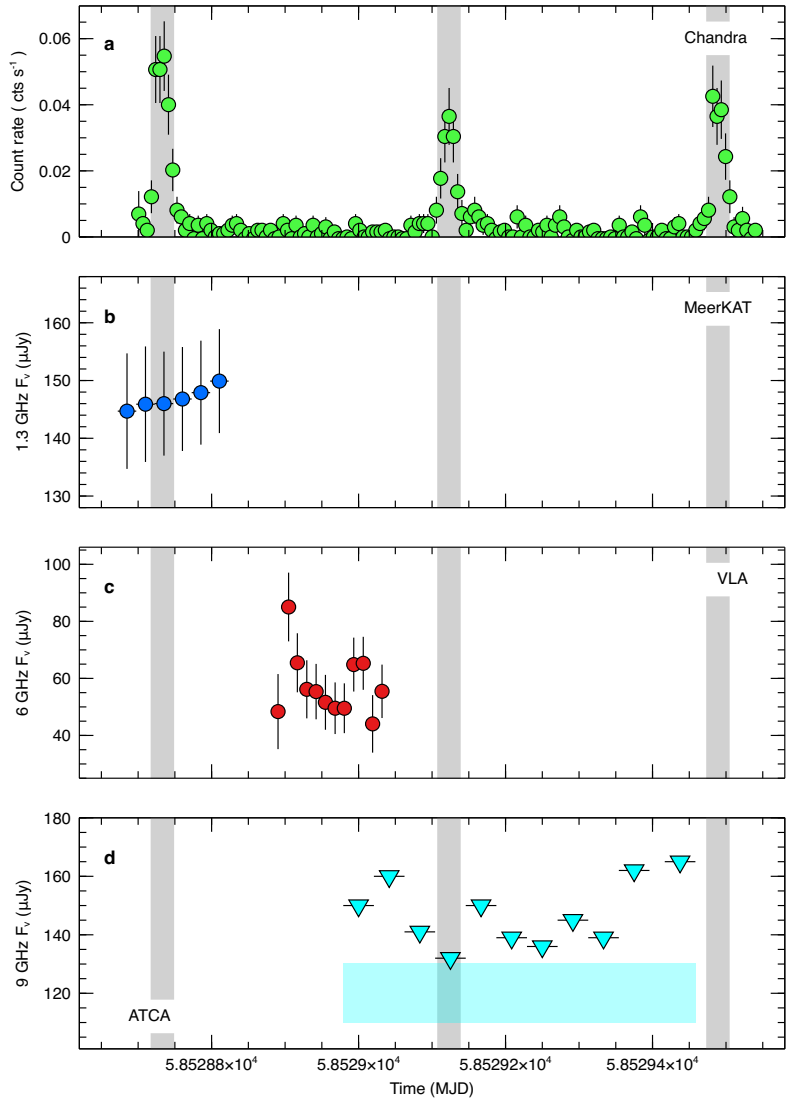

**Extended Data Figure 7 | The 2019 February simultaneous X-ray/radio campaign**
The X-ray 0.4-2 keV Chandra light curve is shown in panel **a**. The remaining panels show the MeerKAT1 (**b**), VLA (**c**), and ATCA2 (**d**) light curves from the simultaneous radio campaign. Notice that the MeerKAT1 and ATCA2 exposures comprise one X-ray QPE each (vertical shaded areas), whereas the VLA observation was performed during X-ray quiescence. No significant radio variability is detected in any of the radio exposures. The ATCA2 data points are all upper limits, and the horizontal shaded area in panel **d** represents the measured time-averaged flux density. We ignore the first data point of the ATCA2 light curve as the source was still very low on the horizon resulting in a highly degraded image. We also point out that the ATCA2 measurements are contaminated by a nearby unresolved radio source detected by the VLA with a flux density of $71 \pm 10$ μJy at 6 GHz. Error bars, including the average flux from ATCA represent the 1-σ confidence intervals. The ATCA data points are instead 3-σ confidence level upper limits.